\documentclass[manuscript]{aastex}

\usepackage[outdir=./]{epstopdf}
\usepackage{graphics,graphicx}
\usepackage{xcolor}

\shorttitle{
}
\shortauthors{Jhan \& Lee}

\def\arcsa#1#2{$#1^{\prime\prime}_{^\textrm{.}}#2$}
\def\cirg#1#2{$#1^{\circ}_{^\textrm{.}}#2$}
\def\ups#1#2{$#1^{s}_{^\textrm{.}}#2$}

\begin{document}

\title{25 AU Angular Resolution Observations of HH 211 with ALMA : Jet
Properties and Shock Structures in SiO, CO, and SO }

%\title{Probing Jet Properties in HH 211 with High Angular Resolution Observations of ALMA
%}

\author{
Kai-Syun Jhan\altaffilmark{1,2} \& Chin-Fei Lee\altaffilmark{2,1}
}

\affil{Graduate Institute of Astronomy and Astrophysics, National Taiwan University, No. 1, Sec. 4, Roosevelt Road, Taipei 10617, Taiwan}

\affil{Academia Sinica Institute of Astronomy and Astrophysics, No. 1, Sec. 4, Roosevelt Road, Taipei 10617, Taiwan}

\altaffiltext{}{E-mail: ksjhan@asiaa.sinica.edu.tw}

\begin{abstract}
 
HH 211 is a highly collimated jet with a chain of knots and a wiggle
structure on both sides of a young Class 0 protostar.  We used two epochs of
Atacama Large Millimeter/submillimeter Array (ALMA) data to study its inner
jet in the CO(J=3-2), SiO(J=8-7), and SO(N$_{J}$=8$_{9}$-7$_{8}$) lines at
$\sim$25 AU resolution.  With these ALMA and previous 2008 Submillimeter
Array (SMA) data, the proper motion of 8 knots within $\sim$250 AU of the
central source is found to be {\bf $\sim$\arcsa{0}{068} per year ($\sim$102 km
s$^{-1}$)}, consistent with previous measurements in the outer jet.  At
$\sim$4 times higher resolution, the reflection-symmetric wiggle can be
still fitted by a previously proposed orbiting jet source model.  Previously
detected continuous structures in the inner jet {\bf are now} resolved,
containing at least 5 sub-knots.  \textbf{These sub-knots are interpreted in
terms of a variation} in the ejection velocity of the jet with a period of
$\sim$ 4.5 years, shorter than that of the outer knots.  In addition,
backward and forward shocks are resolved in a fully-formed knot, BK3, and
signatures of internal working surface and sideways ejection are identified
in Position-Velocity diagrams.  In this knot, low-density SO and CO layers
are surrounded by a high-density SiO layer.

\end{abstract}


\section{Introduction}

Protostellar jets play an important role in the star formation process. In
the theoretical picture, they are believed to be launched and remove excess
angular momentum from the inner part of the accretion disk within 1 AU of
protostars \citep{2014prpl.conf..451F,2020A&ARv..28....1L}.  Two competing
models of jet launching are proposed: the disk-wind model
\citep{2000prpl.conf..759K} and the X-wind model
\citep{2000prpl.conf..789S}.  Both models predict a fast-dense collimated
jet propagating along the rotation axis of the system.
As the inner parts
of the disks where the jets are launched are on scales ($\le$ 1 AU)
{\bf not easy} to be
resolved with current instruments,   we use the jet
properties (e.g., jet morphology and rotation) and knot (shock) structures
observed on large scales to constrain the accretion processes and physical
conditions during the early phase of star formation.  The goal is to study
the mechanism that would remove angular momentum near the protostar.

HH 211 is a well-studied jet consisting of a chain of well-defined knots and is highly collimated with a small-scale wiggle \citep{2010ApJ...713..731L}, and thus it is a good candidate for studying the physical properties of the jet and knots. It is located in the IC 348 complex of Perseus at a distance of 321$\pm$10 pc \citep{2018ApJ...865...73O} and powered by a low-mass \cite[$\sim$ 0.08 M$_{\bigodot}$, after being updated with the new distance from][]{2019ApJ...879..101L} Class 0 protostar \citep{2003MNRAS.346..163F,2005ApJS..156..169F}. It has been
detected in H$_{2}$ \citep{1994ApJ...436L.189M}, as well as SiO, SO, and CO \citep{1999A&A...343..571G,2006ApJ...636L.141H,2006ApJ...636L.137P,2007ApJ...670.1188L}.
The jet consists of a large number of knots on both sides of the central source, with an inclination angle of only $\sim$9 degrees to the plane of sky \citep{2016ApJ...816...32J}. Therefore, HH 211 is a suitable source to study the backward and forward shocks and sideways ejection.

Here, with the Atacama Large Millimeter/submillimeter Array (ALMA), we not
only confirm the previously found wiggle structure, proper motion, and knot
structures, but also study the morphological relationship among the regions
seen in the CO(J=3-2), SiO(J=8-7), and SO(N$_{J}$=8$_{9}$-7$_{8}$) lines. 
SiO and SO are shock tracers
\citep{1992A&A...254..315M,1997A&A...321..293S}, and CO and SO can both
trace jet and outflow.  However, their relationship is still unclear because
of insufficient resolution to separate them.  They may or may not trace
material of different densities ejected from different parts of the disk. 
Now, we have higher resolution {\bf \cite[up to $\sim$\arcsa{0}{06}, which is 4
times better than the previous study in ][]{2009ApJ...699.1584L} }to resolve
the jet and its knots in order to address this issue.

%{\bbf ???? Please check the final resolution you used. Really 0.05" and 5
%times higher???}

\section{Observations}

Observations toward the HH 211 system were obtained with ALMA in 2015 and
2016.  The 2015 observations (cycle 3; Project ID: 2015.1.00024.S) were
already reported in \citet{2018ApJ...863...94L}.  \textbf{They have the same
correlator setup as the 2016 observations (Project ID: 2016.1.00017.S), but
with higher angular resolution to search for jet rotation.}  Here we only
summarize the 2016 observations and compare both observations in Table 1.

HH 211 has been mapped with ALMA at $\sim$ 358 GHz in Band 7 with one
pointing toward the center.  A C40-7 array configuration was used with 43
antennas.  Two executions were carried out in 2016 in Cycle 4 both on
October 10.  The projected baselines were 13-3145 m.  We had 5 spectral
windows in the correlator setup, with four for molecular lines at a velocity
resolution of $\sim$0.212 km s$^{-1}$ and one for continuum at a velocity
resolution of $\sim$0.848 km s$^{-1}$.  The CASA package was used to
calibrate the data for passband, flux, and gain.  The quasars J0238+1636,
J0237+2848, and J0336+3218 were used as the flux, bandpass, and gain
calibrator, respectively.  \textbf{A phase-only self-calibration derived
from the continuum was also applied to the $uv$ data in order to improve the
image fidelity and reduce the sidelobes.  Since the $uv$ data in long
baselines have a signal to noise ratio lower than 3, we only used the
continuum data with $uv$ distance $< 1000$ m to derive the solution for the
self-calibration and applied the self-calibration to the data with $uv$
distance $< 1000$ m.}

The CASA package was used to image the data. We used CONCAT to combine both
the 2015 and 2016 data, and then CLEANed the combined data {\bf to generate
various line intensity maps with a velocity resolution of $\sim$ 1 km
s$^{-1}$.  In order to better show the structure of the jet and outflow
shells, we used a robust weighting factor of 2 for the visibility to
generate the maps.  The resulting CO(J=3-2), SiO(J=8-7), and
SO(N$_{J}$=8$_{9}$-7$_{8}$) maps have a noise level of $\sim$ 1.1, 1.1, and
1.0 mJy beam$^{-1}$, respectively.  The synthesized beam has a size of
\arcsa{0}{139} $\times$ \arcsa{0}{083} at a position angle (P.A.) of
\cirg{-9}{8} for SiO, \arcsa{0}{129} $\times$ \arcsa{0}{069} at a P.A.  of
\cirg{-8}{9} for CO, and \arcsa{0}{13} $\times$ \arcsa{0}{069} at a P.A.  of
\cirg{-7}{8} for SO.  We also used a robust weighting factor of 0.5 for the
visibility to generate higher-resolution maps to search for jet rotation and to
study shock structures in knot BK3.  The resulting CO, SiO, and SO maps
have a noise level of $\sim$ 1.3, 1.2, and 1.2 mJy beam$^{-1}$,
respectively.  The synthesized beam has a size of \arcsa{0}{109} $\times$
\arcsa{0}{059} at a position angle (P.A.) of \cirg{-12}{2} for SiO,
\arcsa{0}{095} $\times$ \arcsa{0}{057} at a P.A.  of \cirg{-13}{7} for CO,
and \arcsa{0}{095} $\times$ \arcsa{0}{057} at a P.A.  of \cirg{-13}{0} for
SO.  In these maps, the velocities are in the LSR system.}

%{\bbf ??? Please update and complete the above paragraph????}

\section{Results}

In HH 211, the systemic velocity is $V_{sys}$ = 9.2 $\pm$ 0.1 km s$^{-1}$ \citep{1999A&A...343..571G}. In this system, the jet has a position angle of \cirg{116}{1} with the southeastern (SE) component tilted toward us \citep{1999A&A...343..571G}. 
The continuum emission peak is at the position of $\alpha_{(2000)}$ = 3$^{h}$43$^{m}$\ups{56}{8054} and $\delta_{(2000)}$ = 32$^{\circ}$00$^{\prime}$\arcsa{50}{189} \citep{2018ApJ...863...94L}.

\subsection{Jet Morphology}

\begin{figure}
\epsscale{1}
\plotone{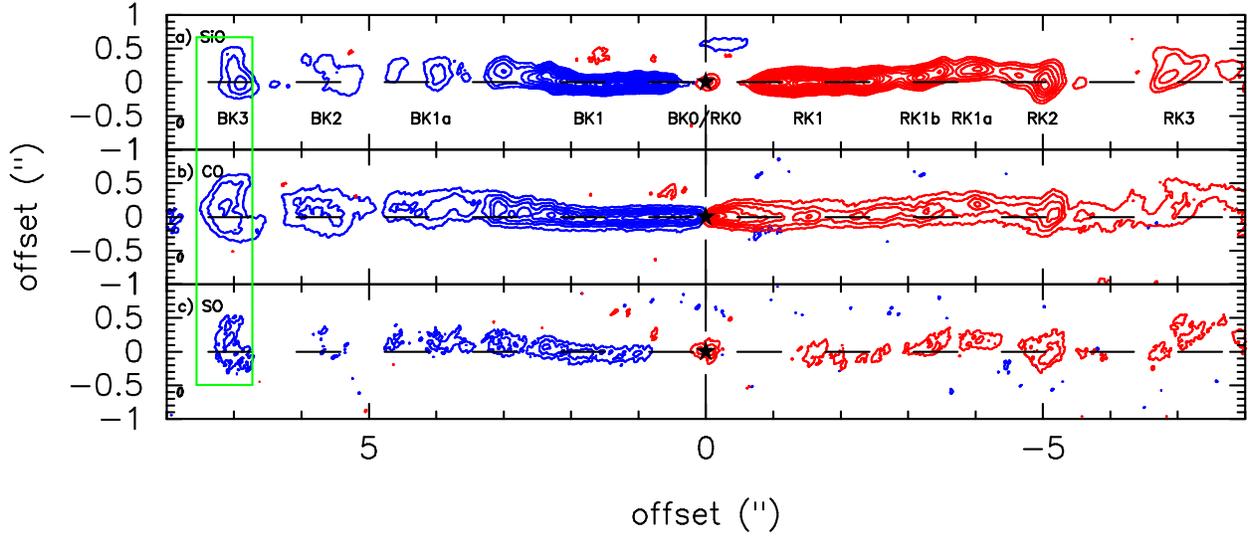}
\centering
\caption{ (a) SiO emission map. The blue and the red contours are the blueshifted (\textbf{-24.8$\le$V$\le$9.2 km s$^{-1}$) and the redshifted (43.2$\ge$V$\ge$9.2 km s$^{-1}$}) emissions of the SiO. The contour levels start at 3 $\sigma$ with a step of 6 $\sigma$, and $\sigma$ is $\sim$0.2 and 0.19 mJy beam$^{-1}$ for the blueshifted and the redshifted sides, respectively. (b) CO emission map. The blue and the red contours are the total high velocity blueshifted (\textbf{-14.8$\le$V$\le$1.2 km s$^{-1}$) and the total high velocity redshifted (\textbf{35.2$\ge$V$\ge$19.2 km s$^{-1}$}}) emissions of the CO, and the contour levels start at 3 $\sigma$ with a step of 6 $\sigma$, and $\sigma$ is $\sim$0.3 mJy beam$^{-1}$. (c) SO emission map. The blue and the red contours are the blueshifted (\textbf{-19.8$\le$V$\le$9.2 km s$^{-1}$) and the redshifted (41.2$\ge$V$\ge$9.2 km s$^{-1}$}) emissions of the SO, and the contour levels start at 3 $\sigma$ with a step of 6 $\sigma$, and $\sigma$ is $\sim$0.21 and 0.19 mJy beam$^{-1}$, respectively.
}
\label{siocoso}
\end{figure}

SiO(J=8-7), CO(J=3-2), and SO(N$_{J}$=8$_{9}$-7$_{8}$) are detected on both
sides of the central source, and their emission intensity maps are shown in
Figure \ref{siocoso}(a), (b), and (c), respectively.  As seen before
\citep{2009ApJ...699.1584L}, the jet in these lines is highly collimated
with a small wiggle and consists of a chain of knots, even though the SO
emission is relatively weak.  Also, SiO, SO, and CO emissions trace a bow
\textbf{shock} structure for knot BK3 in the jet (green box in Figure
\ref{siocoso}).

\begin{figure}
\epsscale{1}
\plotone{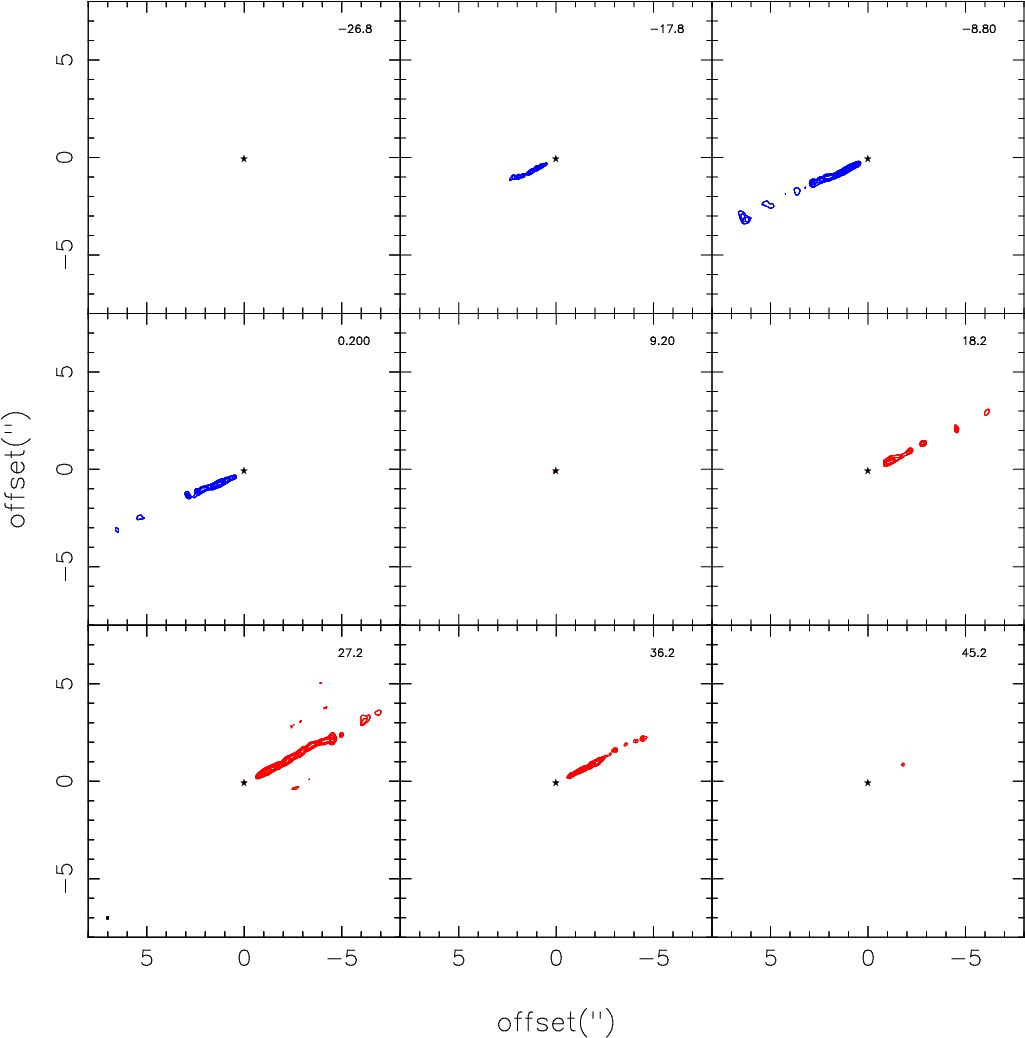}
\centering
\caption{ SiO channel maps from -26.8 km s$^{-1}$ to 45.2 km s$^{-1}$ with a step of 9 km s$^{-1}$ and a velocity resolution of 9 km s$^{-1}$. The contour levels start at 10 $\sigma$ with a step of 20 $\sigma$, and $\sigma$ is $\sim$0.38 mJy beam$^{-1}$. The number in the top-right corner is the radial velocity of the maps in km s$^{-1}$.   \label{siochamap}}
\end{figure}

\begin{figure}
\epsscale{1}
\plotone{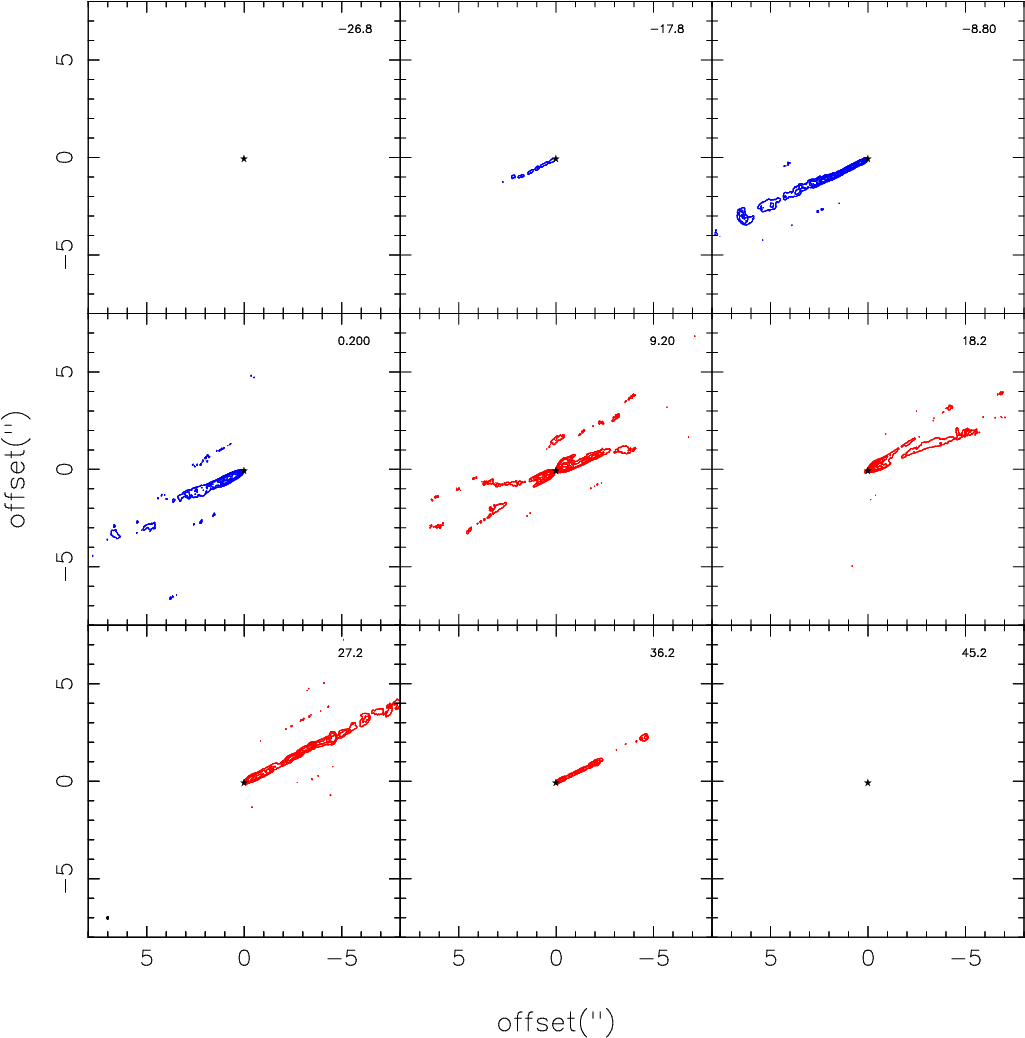}
\centering
\caption{ CO channel maps from -26.8 km s$^{-1}$ to 45.2 km s$^{-1}$ with a step of 9 km s$^{-1}$ and a velocity resolution of 9 km s$^{-1}$. The contour levels start at 5 $\sigma$ with a step of 10 $\sigma$, and $\sigma$ is $\sim$0.38 mJy beam$^{-1}$. The number in the top-right corner is the radial velocity of the maps in km s$^{-1}$.   \label{cochamap}}
\end{figure}

\begin{figure}
\epsscale{1}
\plotone{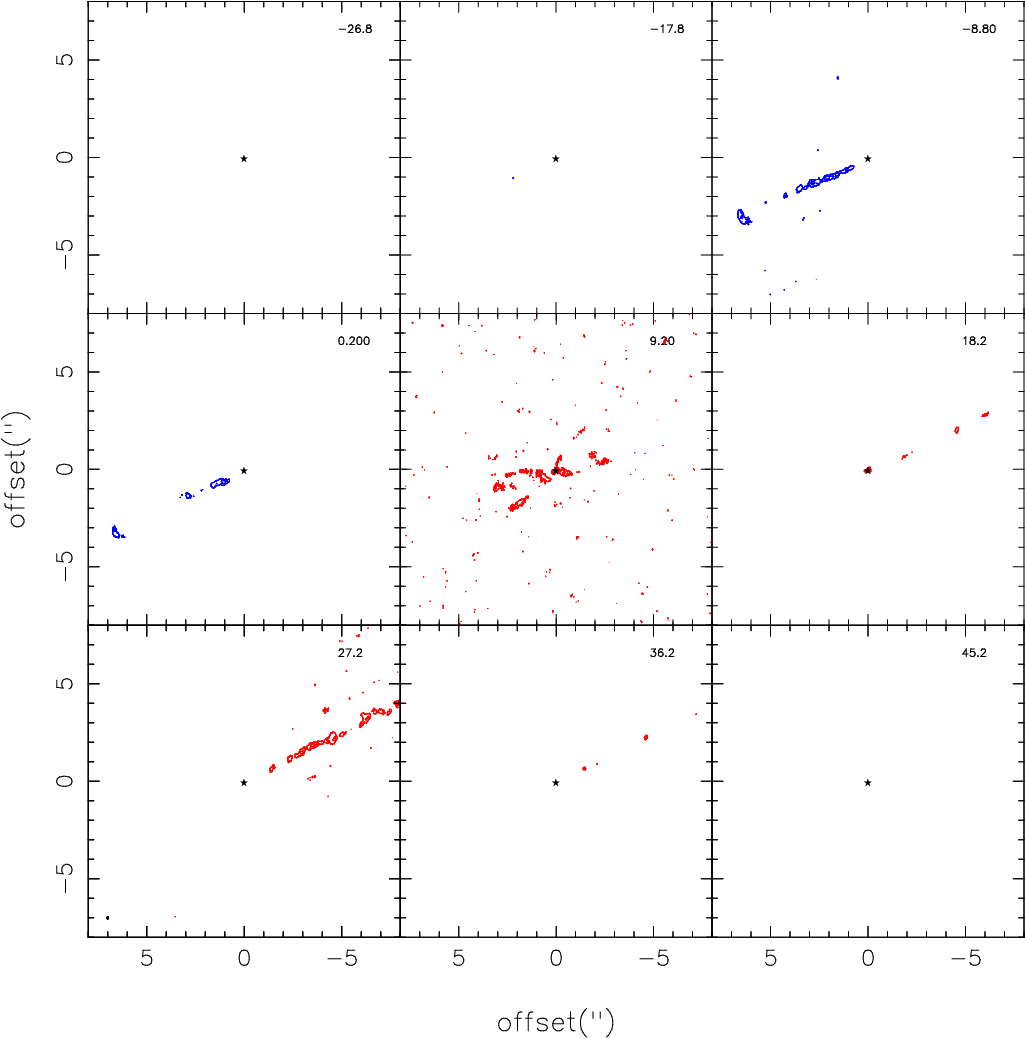}
\centering
\caption{ SO channel maps from -26.8 km s$^{-1}$ to 45.2 km s$^{-1}$ with a step of 9 km s$^{-1}$ and a velocity resolution of 9 km s$^{-1}$. The contour levels start at 5 $\sigma$ with a step of 10 $\sigma$, and $\sigma$ is $\sim$0.37 mJy beam$^{-1}$. The number in the top-right corner is the radial velocity of the maps in km s$^{-1}$.   \label{sochamap}}
\end{figure}

In channel maps of these emissions (Figure \ref{siochamap} for SiO, Figure
\ref{cochamap} for CO, and Figure \ref{sochamap} for SO), the jet appears
more and more collimated going from low to high velocity.  In CO, the jet is
seen at high velocity (V$_{LSR}$ $\le$ 0.2 km s$^{-1}$ and V$_{LSR}$ $\ge$
18.2 km s$^{-1}$), and outflow shells are seen at low velocity (0.2 km
s$^{-1}$ $\le$ V$_{LSR}$ $\le$ 18.2 km s$^{-1}$), as seen before
\citep{1999A&A...343..571G,2007ApJ...670.1188L}.  This distribution is also
seen in other objects \citep{2009A&A...495..169S,2015ApJ...805..186L}.  SO
traces the jet as SiO and CO does, and shows evidence of shell-like
structure at V$_{LSR}$=9.2 km s$^{-1}$ within 2$^{\prime\prime}$ of the
central source tracing the base of the outflow.  All three lines can also
trace bow \textbf{shock} structures, e.g., at velocity $\sim$-8.8 and 0.2 km
s$^{-1}$ on the blueshifted side, for example.

\subsection{Proper Motion}

\begin{figure}
\epsscale{1}
\plotone{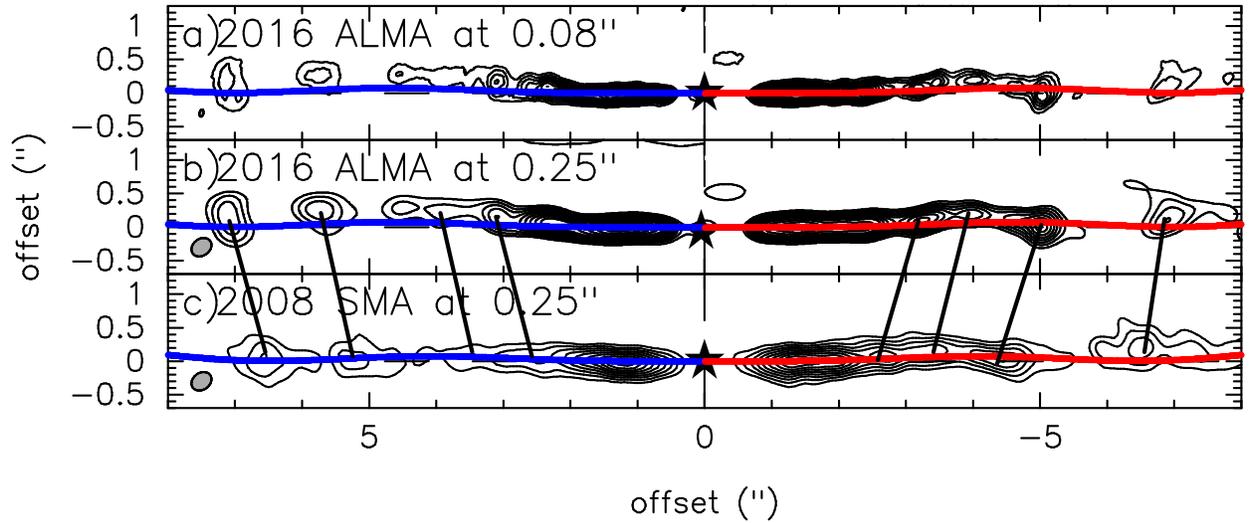}
\centering
\caption{ \textbf{The comparison between the observed jet structure and the orbiting jet source model. The observed jets here are rotated by \cirg{26}{1} clockwise.} (a) SiO map at \arcsa{0}{08} resolution of the 2016 ALMA observation before convolution. The contour levels start at 3 $\sigma$ with a step of 6 $\sigma$, and $\sigma$ is $\sim$0.17 mJy beam$^{-1}$. (b) SiO map at \arcsa{0}{25} resolution of the 2016 ALMA observation after convolution. The contour levels start at 3 $\sigma$ with a step of 30 $\sigma$, and $\sigma$ is $\sim$1.6 mJy beam$^{-1}$. (c) SiO map at \arcsa{0}{25} resolution of the 2008 SMA observation. The contour levels start at 3 $\sigma$ with a step of 3 $\sigma$, and $\sigma$ is $\sim$1.98 mJy beam$^{-1}$.
}
\label{convol}
\end{figure}

In 2008, we observed HH 211 at a resolution of \arcsa{0}{25} with  the
Submillimeter Array (SMA) \cite[][Figure
\ref{convol}(c)]{2009ApJ...699.1584L}.  Therefore, proper motion can be
estimated by measuring the position shift of the knots from the 2008 SMA
observation to the 2016 ALMA observation (Figure \ref{convol}(a) and (b)). 
The time interval is 7.72 years.  We first degraded the 2016 ALMA map from
\arcsa{0}{08} to \arcsa{0}{25} resolution and aligned the maps with the
continuum peaks.  In these two epochs, 8 knots within
$\sim$8$^{\prime\prime}$ of the central source, with 4 on each side of the
source, are used for proper motion measurement.  We measured the peak
positions of these knots (Figure \ref{convol}) and calculated the position
shift of the knots.  The error bar of position shift is assumed to be one
third of the beam size along the jet axis.  The proper motion is then
estimated to be {\bf$\sim$\arcsa{0}{068} $\pm$ \arcsa{0}{01} per year.  With
the new distance of 321 pc, the transverse velocity is $\sim$102 $\pm$ 16 km
s$^{-1}$.} Thus, the inclination angle is refined to be $\sim$11 degrees,
using a mean radial velocity of $\sim$20 km s$^{-1}$ estimated later in
Section 3.5.

\begin{figure}
\epsscale{1}
\plotone{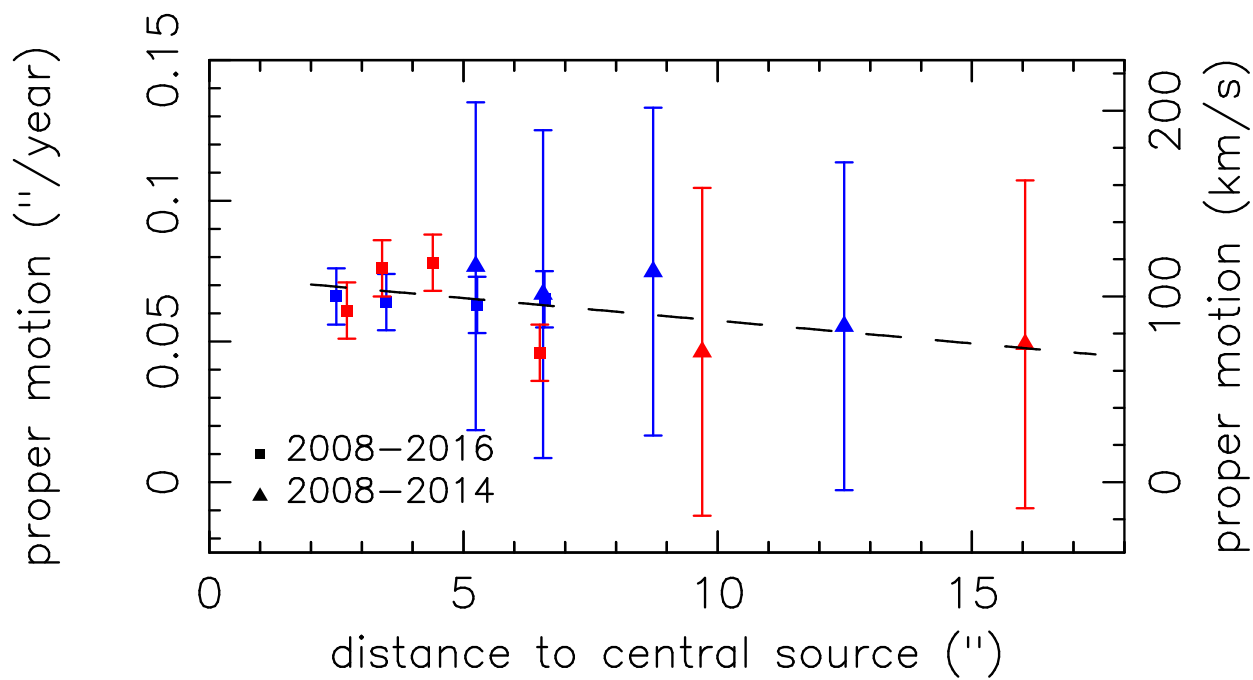}
\centering
\caption{ Relation of the proper motion and distance of the knots to the central source. The square symbols represent the proper motion measurements using 2016 ALMA results and 2008 SMA results. The triangle symbols represent the proper motion measurements using 2008, 2010, and 2013 SMA results, i.e., the measurements of \cite{2016ApJ...816...32J} without 2004 SMA results. The red color indicates the measurements of the knots on the redshifted side, and the blue color indicates the measurements of the knots on the blueshifted side. The dashed line is a linear fit of proper motion and distance of all knots in both measurements.  \label{fitv}}
\end{figure}

We then compared our result with the previous SMA result reported in \citet{2016ApJ...816...32J} for the outer parts of the jet at a distance from 5$^{\prime\prime}$ to 16$^{\prime\prime}$. We first refined the previous SMA measurements by taking out the 2004-epoch data due to its poorest resolution ($\sim$\arcsa{1}{2}) and poorest UV coverage compared to those in other epochs. The refined proper motion (triangle symbols in Figure \ref{fitv}) is 30 \% lower than that measured before, but is closer to our current ALMA measurement for the inner jet, as shown in Figure \ref{fitv}. The figure shows that the velocity of the jet is roughly constant at a distance from 1000 to 5000 AU (3$^{\prime\prime}$ to 16$^{\prime\prime}$) within the error bar, although there could be a slight decrease in velocity with distance, as shown by the linear fit.

%{\bbf ????? The right ticks in Figure 6 are wrong. Please correct them. ???}

%Then we plotted them in Figure \ref{fitv} with the distance to the central source. The high resolution proper motion measurement contains knots in the inner part of the jet (distance to the source $<$ 7$^{\prime\prime}$), and the low resolution measurement contains knots in the outer part of the jet (distance to the source $>$ 5$^{\prime\prime}$).These two measurements are consistent within the error bar, thus

%However, if we included all the data in these measurements and fit the velocity with the distance to the central source with linear equation (dashed line in Figure \ref{fitv}), this fit shows that the jet does slow down with distance from the central source. This is probably due to the difference of the error bars in two measurements: The high-resolution measurement (with lower error) only measure the proper motion of knots within 7$^{\prime\prime}$, and the low-resolution (with higher error) measure the proper motion of knots beyond 5$^{\prime\prime}$. Therefore, we could not tell if the slow-down problem is real or due to the resolution. More observation with high resolution (with lower error) contain both regions is needed to solve this problem.

\subsection{Wiggle Structure}

A reflection-symmetric wiggle has been detected in the jet and modeled with
an orbiting jet source model in \citet{2010ApJ...713..731L}.  The same model
with the same parameters were later used to fit the wiggle in 3 other epochs
(2004, 2010, and 2013) of the SMA data of the SiO line
\citep{2016ApJ...816...32J}.  Now, we can check whether the model can still
fit the same wiggle in our ALMA observation in the inner part of the jet at
higher resolution.  In our mapped region, the jet shows one to two cycles of
wiggle, as shown in Figure \ref{convol}(a).  After adjusted for proper
motion, the previous best-fit parameters still roughly fit the pattern of
the wiggle but with some deviations, especially in the regions, from
3$^{\prime\prime}$ to 6$^{\prime\prime}$ and from -3$^{\prime\prime}$ to
-4$^{\prime\prime}$ where {\bf the peaks of the knots deviate to the north from the
model.}  We will discuss the deviations in more detail in
Section 4.1.

\subsection{Kinematics Perpendicular to the Jet Axis}

\begin{figure}
\epsscale{0.6}
\plotone{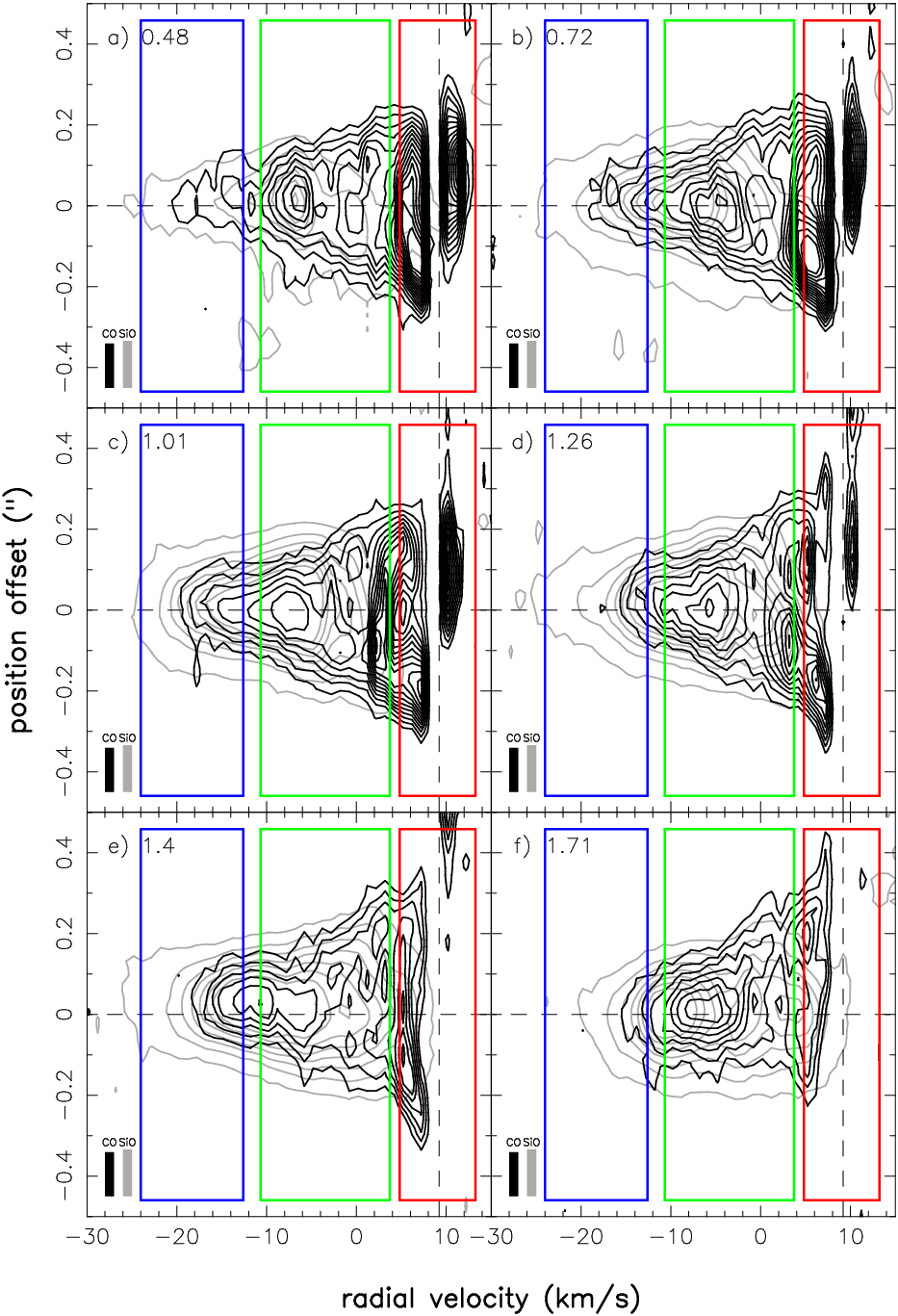}
\centering
\caption{ SiO (gray contour) and CO (black contour) PV diagrams of the blueshifted knots \textbf{along cuts} perpendicular to the jet axis. For SiO, the contour levels start at 3 $\sigma$ with a step of 15 $\sigma$, and $\sigma$ is $\sim$0.95 mJy beam$^{-1}$. For CO, the contour levels start at 3 $\sigma$ with a step of 3 $\sigma$, and $\sigma$ is $\sim$1.1 mJy beam$^{-1}$. The horizontal dashed lines indicates the zero position of the cuts, and the vertical dashed lines indicates systemic velocity. The numbers on the upper left corner indicate the distance from the source to the zero position of the cuts. The blue, green, and red boxes denote the high, middle, and low velocity components, respectively.
}
\label{pvbkk}
\end{figure}

\begin{figure}
\epsscale{0.5}
\plotone{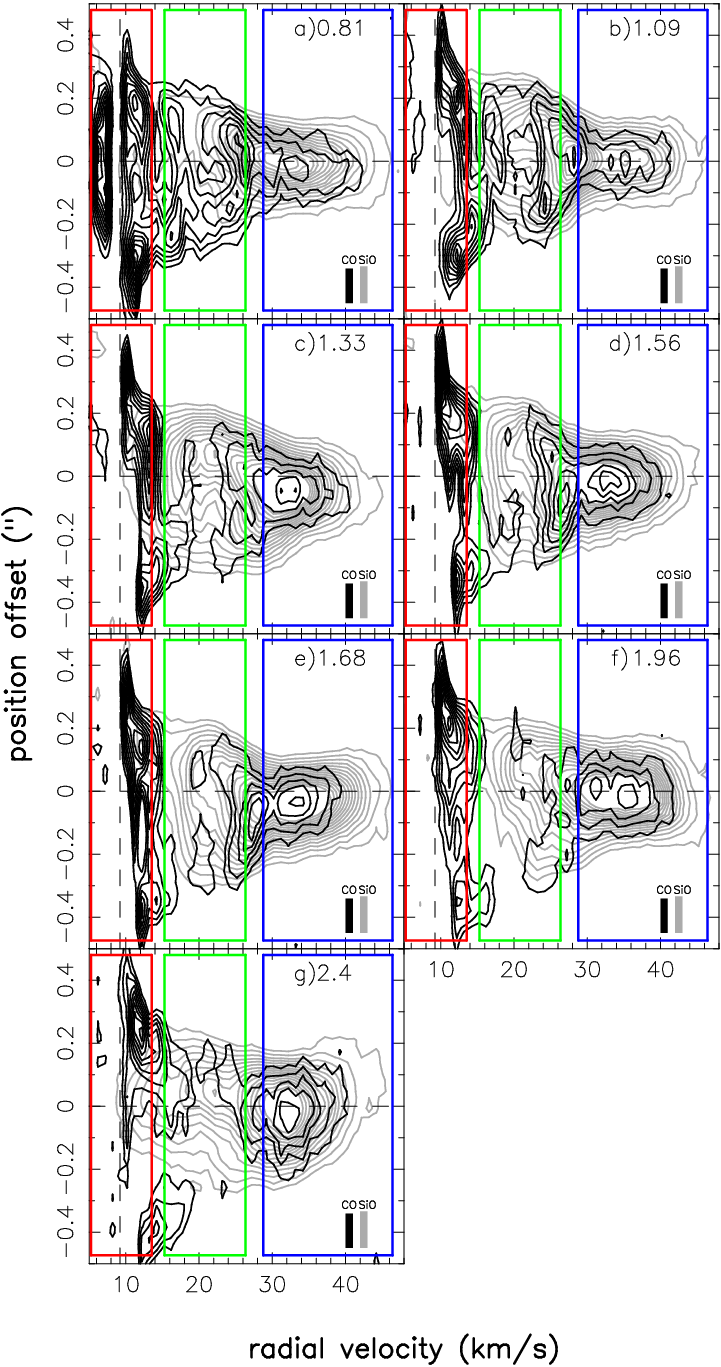}
\centering
\caption{ SiO (gray contour) and CO (black contour) PV diagrams of the redshifted knots \textbf{along cuts} perpendicular to the jet axis. For SiO, the contour levels start at 3 $\sigma$ with a step of 6 $\sigma$, and $\sigma$ is $\sim$0.95 mJy beam$^{-1}$. For CO, the contour levels start at 3 $\sigma$ with a step of 3 $\sigma$, and $\sigma$ is $\sim$1.1 mJy beam$^{-1}$. The horizontal dashed lines indicates the zero position of the cuts, and the vertical dashed lines indicates systemic velocity. The numbers on the upper right corner indicate the distance from the source to the zero position of the cuts. The blue, green, and red boxes denote the high, middle, and low velocity components, respectively.
}
\label{pvrkk}
\end{figure}

To study the jet kinematics perpendicular to the jet axis, we presented the
position-velocity (PV) diagrams \textbf{along cuts} perpendicular to the
jet axis in SiO and CO emissions (Figure \ref{pvbkk} to Figure \ref{pvrkk})
centered at the SiO emission peaks which have a distance of \arcsa{0}{5} to
2$^{\prime\prime}$ from the central source.  We identified three velocity
components in the diagrams, as marked by boxes of different colors in the
figures: a low-velocity component in red boxes with a wide spatial range, a
middle-velocity triangular component in green boxes, and a high-velocity
linear component in blue boxes.  The low-velocity component is seen mainly
in CO, tracing the outflow shell as discussed in Section 3.1.  The
middle-velocity triangular structure appears in CO and SiO.  The
high-velocity linear structure is also seen in CO and SiO with almost the
same velocity range in these two lines.  This high-velocity structure traces
the jet itself, as seen in the channel maps.  However, the highest velocity
of the jet gradually slows down with increasing distances \cite[as can be
seen in blue boxes at different distance, or see Figure 8
in][]{2004ApJ...606..483L}, and then the high-velocity parts merge
with the triangular components in some of the diagrams at large distance.

\begin{figure}
\epsscale{0.8}
\plotone{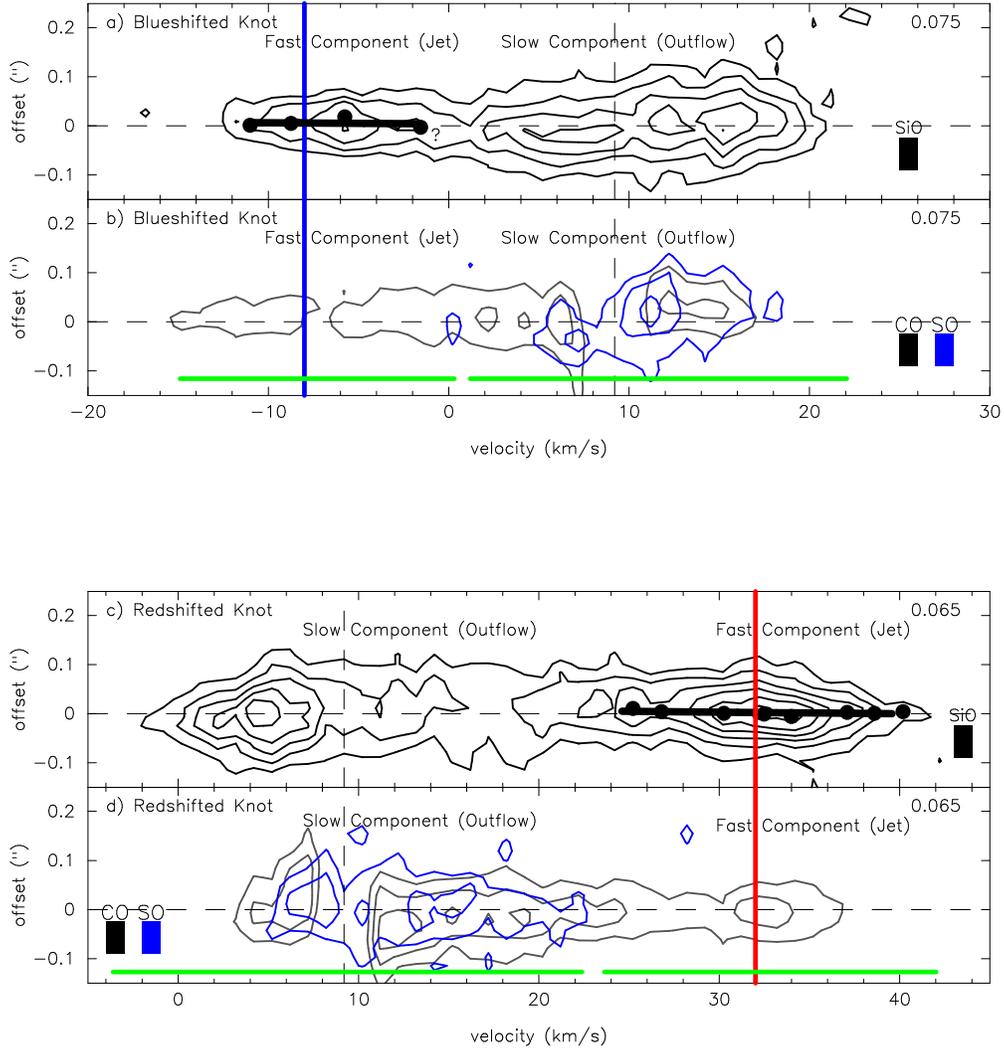}
\centering
\caption{ (a)(c) SiO PV diagrams of knots BK0 and RK0 \textbf{along cuts} perpendicular to the jet axis. The contour levels start at 3 $\sigma$ with a step of 3 $\sigma$, and $\sigma$ is $\sim$2.0 mJy beam$^{-1}$.  (b)(d) CO(gray contour) and SO (blue contour) PV diagrams of knots BK0 and RK0 \textbf{along cuts} perpendicular to the jet axis. The contour levels start at 3 $\sigma$ with a step of 3 $\sigma$ for CO and 2  $\sigma$ with a step of 2 $\sigma$ for SO. $\sigma$ is $\sim$2.0 mJy beam$^{-1}$ and $\sim$1.9 mJy beam$^{-1}$ for CO and SO, respectively.  The blue vertical line marks the mean velocity of the jet on the blueshifted side, which is -8 km s$^{-1}$. The red vertical line marks the mean velocity of the jet on the redshifted side, which is 32 km s$^{-1}$. The round symbols dot the peak positions, and the thick line roughly connects those peaks. The numbers on the upper right corner indicate the distance from the source to the zero position of the cuts. The green segments denote the velocity range of the slow component defined by the SO emission.
%{\bbf please align and place the labels properly. Also larger fonts???}
}
\label{pv1}
\end{figure}

The first pair of knots (knots BK0 and RK0) in SiO (Figure \ref{siocoso}(a))
appear within \arcsa{0}{1} of the source and are separated from the other
knots downstream.  Figure \ref{pv1} shows their PV diagrams \textbf{along
cuts} perpendicular to the jet axis in SiO, CO, and SO.  SiO and CO emission
appears roughly in the same velocity range, but SO emission appears only at
low velocity within $\sim$10 km s$^{-1}$ of the systemic velocity. 
Therefore, a slow component is defined with the SO emission using the SO
velocity range (from $\sim$1 to 20 km s$^{-1}$ in the blueshifted knot and
from $\sim$-4 to 23 km s$^{-1}$ in the redshifted knot).  This slow
component in SO has been reported in \citet{2018ApJ...863...94L}.  At higher
angular resolution in SO, this slow outflow was found to be rotating, with a
velocity gradient across the jet axis centered at the systemic velocity
\cite[see Figure 4 in][]{2018ApJ...863...94L}, and thus can trace a disk
wind.  This velocity gradient can also be roughly seen here in SO.

%{\bbf ????The slow component defined by SO has different velocity ranges now? %Is it correct?
%Could you please check or plot lower contour levels????}
%A velocity gradient is seen on the blueshifted side across 12 km s$^{-1}$, with the velocity sense the same as the rotation in the envelope seen in HCO$^{+}$ \citep{2009ApJ...699.1584L}, and thus may trace a rotating outflow, as suggested in \citet{2018ApJ...863...94L}. However, on the redshifted side, the slow component ($\sim$5 km s$^{-1}$) does not show such a velocity gradient probably because of an insufficient resolution. 

%These slow components in three line emissions may trace outflow, and we can see a velocity gradient on the blueshifted side around 12 km s$^{-1}$. It shows a rotating sense that the material rotates in the same direction as HCO$^{+}$ and SO gas observed before \citep{2009ApJ...699.1584L, 2018ApJ...863...94L}. However, on the redshifted side, the slow component ($\sim$5 km s$^{-1}$) does not show such a velocity gradient probably because of insufficient resolution. 

In addition, SiO and CO also trace a fast component ($\sim$-15 to 0 km
s$^{-1}$ in the blueshifted knot and $\sim$24 to 42 km s$^{-1}$ in the
redshifted knot), {\bf allowing us to search for jet rotation.} This fast
component, with the central velocity close to the velocity of the jet (as
discussed in Section 3.5), traces the jet itself.  {\bf However, no clear
velocity gradient can be seen across these knots at our current resolution.}

\subsection{Kinematics along the Jet Axis}

\begin{figure}
\epsscale{1}
\plotone{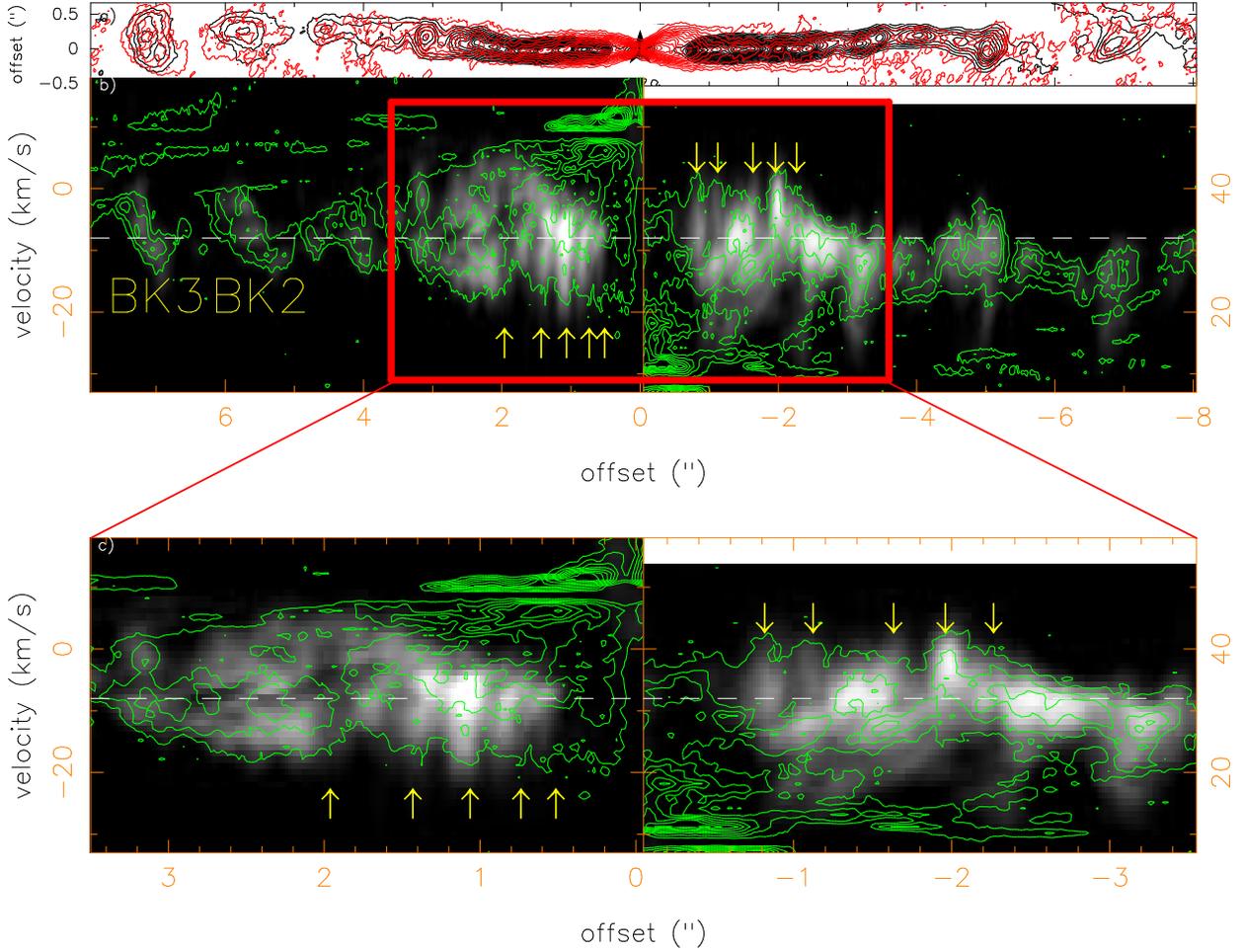}
\centering
\caption{ (a) SiO (black contour) and CO (red contour) intensity maps. For SiO, the contour levels start at 3 $\sigma$ with a step of 6 $\sigma$, and $\sigma$ is $\sim$0.2 mJy beam$^{-1}$. For CO, the contour levels start at 3 $\sigma$ with a step of 6 $\sigma$, and $\sigma$ is $\sim$0.1 mJy beam$^{-1}$. The asterisk symbol denotes the position of central source. (b)(c) SiO (gray color) and CO (green contour) PV diagram. For CO, the contour levels start at 3 $\sigma$ with a step of 6 $\sigma$, and $\sigma$ is $\sim$0.95 mJy beam$^{-1}$. The white horizontal dashed lines mark the mean velocities, which are -8 km s$^{-1}$ on the blueshifted side and 32 km s$^{-1}$ on the redshifted side. The yellow arrows denote the positions of the sub-knots.
}
\label{pvall}
\end{figure}

To study the kinematics along the jet axis, we present PV diagrams cut along
the jet axis in CO and SiO (Figure \ref{pvall}).  The PV structures in CO
and SiO are roughly the same.  In addition, the PV structures on the
blueshifted side and the redshifted side are also roughly the same.  {\bf Note
that in Figure \ref{pvall}(a), the roughly linear structure seen in CO (red
contours) on the right side of the jet (roughly at $x$ offsets
from -\arcsa{2}{6} to -\arcsa{4}{0})
that appears on the other side of the main wiggle 
is the outflow structure around the jet.}
%{\bbf Nick, expand Figure 11a to have y-offset of +-0.5. .....}

%The PV structure on the blueshifted side is knotty with the same patterns which means that the ejection velocity varies with time \citep{2016ApJ...816...32J}. On the other hand, the PV structure on the redshifted side is continuously distributed within 8$^{\prime\prime}$ which means that the source continuously ejects material, and this continuously distribution of emission also appears in CO (Figure \ref{siocoso}(b) and Figure \ref{pvall}(a)). 

The knotty structures (knots BK1a, BK2, and BK3) on the blueshifted side are
clear.  Previously, \citet{2016ApJ...816...32J} derived the distance for an
internal shock to fully form, which is{\bf $\sim v_j \cdot \frac{v_j}{\Delta
v/(P/2)} \sim$ \arcsa{2}{6}} updated for the new proper motion measurement,
where {\bf$v_j \sim$102 km s$^{-1}$} is the jet velocity, $\Delta v \sim$30 km
s$^{-1}$ is the amplitude of the velocity variation (based on the velocity
range of knot BK1), and $P \sim$20 years is the period of the velocity
variation (based on the mean interknot spacing and the proper motion). 
Thus, these knot structures should be produced by fully formed internal
shocks.  Their PV structures are all linear with the slow material ahead of
the fast material.  This indicates that, when the fast material catches up
with the slow material, it will form an internal working surface between
them \cite[see Figure 1 in][]{1997A&A...318..595S}.  In Figure \ref{bk32},
we can see that the PV structure of the knot BK3$ $ shows several peaks.  As
discussed later, these peaks likely point out the positions of the backward
(peaks at higher velocity) and forward (peaks at lower velocity) shock, and
the internal working surface (point at middle velocity) formed in the knot
\citep{1990ApJ...364..601R,1993ApJ...413..210S,1997A&A...318..595S,2004ApJ...606..483L,2016ApJ...816...32J}.

Within a distance of{\bf \arcsa{2}{6}}, sub-knots of knots BK1 and RK1 can be
seen on both blueshifted and redshifted sides in the intensity maps (Figure
\ref{pvall}a).  Although they are hard to be identified in the PV
diagrams, we can still see 4 to 5 cycles of velocity variations, with a
variation range of about 25 km s$^{-1}$ to 30 km s$^{-1}$ for these
sub-knots, consistent with the result in \citet{2009ApJ...699.1584L}.

Figure \ref{pvall}(c) zooms into the continuous structures, showing the sub-knots structures in them. The distance between two sub-knots is about \arcsa{0}{3}, thus the period of the sub-knot ejection is about 4.5 years, using a jet velocity of \textbf{102 km s$^{-1}$}. Therefore, using the same formula, the distance for an internal shock to fully form in the sub-knots is $\sim$\arcsa{0}{6}. This is consistent with our observations, which show CO but no SiO emission between \arcsa{0}{1} to \arcsa{0}{5} and -\arcsa{0}{1} to -\arcsa{0}{5}, because the shocks have not formed yet, and thus no SiO emission can be detected. This result is also aligned with a previous simulation that suggested that a presence of a velocity variation with a smaller period is needed to produce the bright emission in the continuous SiO structure near the source \citep{2016MNRAS.460.1829M}. In summary, the PV diagrams suggest at least two different periods of variations: one to form the sub-knot regions near the central source, and the other to form the knotty structure further downstream the jet.

\subsection{Column Density and SiO and SO Abundances in Knot BK3}

The column density of knot BK3 can be estimated from the CO emission.
Assuming that the CO emission is LTE and optically thin, and the abundance
of CO relative to H$_{2}$ is $\sim$4$\times$10$^{-4}$, as if the CO gas is
formed via gas-phase reactions in an initially atomic jet
\citep{1991ApJ...373..254G}, the column density of the knot in {\bf H$_{2}$ is N
$\sim$ 5.6$\times$10$^{20}$ cm$^{-2}$.  Since the brightness temperature is
$\sim$90 K, we assume the excitation temperature to be 200 K, as adopted
in \citet{2010ApJ...713..731L}.
This excitation temperature is 
slightly lower than the kinetic temperature derived from the CO
emission in much higher transitions in the far-infrared, which
was found to be $\gtrsim$ 250 K \citep{Giannini2001}.}
Then, the (two-sided) mass-loss rate of the jet can be given by {\bf $\dot{M}_{j}
\sim 2v_{j}m_{H_{2}}Nb$ $\sim$1.13$\times$10$^{-6}$ M$_{\bigodot}$ yr$^{-1}$},
where N, $m_{H_{2}}$, and b are the column density of the jet, the mass of
H$_{2}$ molecule, and the beam size perpendicular to the jet axis (assuming
that the size of the jet beam is $\sim$\arcsa{0}{4}, $\sim$128 AU, discussed
in Section 4.2), respectively.  This mass-loss rate is consistent with the
previous measurement \citep{2010ApJ...713..731L} after adjusted for the new
proper motion.

SiO and SO abundances in knot BK3 can be derived from their column densities
and that of H$_{2}$.  {\bf The kinetic temperature was found to be 300$-$500
K in knots BK1 and RK1 \textbf{\citep{2006ApJ...636L.141H}}.  Since knot BK3
is further away and expanding, its excitation temperature should be lower
and is thus assumed to be 300 K.} This excitation temperature of SiO and
SO is higher than that of CO because both emissions seem to trace stronger
shocks than the CO emission does.  Assuming that both emissions are
optically thin, the SiO and SO column density are {\bf $\sim$7.9$\times$10$^{14}$
cm$^{-2}$ and $\sim$1.0$\times$10$^{14}$ cm$^{-2}$, resulting in SiO and SO
abundances of $\sim$1.4$\times$10$^{-6}$ and $\sim$1.8$\times$10$^{-7}$,
respectively.}  Notice that the abundance of SiO in quiescent molecular
clouds is $\le$1$\times$10$^{-11}$ \citep{1989ApJ...343..201Z}, and the
abundance of SO in cold quiescent clouds (cores) is $\sim$5$\times$10$^{-9}$
\citep{1998FaDi..109..205O}.  Therefore, both SiO and SO are highly enhanced
by shocks in the knot.

\section{Discussion}

\subsection{Wiggle Structure and Protobinary?}

As mentioned in Section 3.3, the orbiting jet source model could still fit
our current observation with some deviations.  Since this model can also fit
the wiggle structure further out in another 3 epochs over a period of 9
years, we believe that the model is roughly correct, but needs some
refinement.  Comparing our observational results in Figure \ref{convol}(b)
to 2008 SMA observations in Figure \ref{convol}(c) at the same resolution,
we found that the \textbf{peaks} of the knots (knot BK1a, BK2, BK3) in our
ALMA observations deviate to \textbf{the north from the model, with little
flux detected on the south.  This deviation does not appear in the SMA
observations, in which the peaks of the knots are aligned with the model}.

One possible reason is that our ALMA observation results suffer from missing
flux, so some flux of the knots is missing.  The maximum recoverable size
(MRS) is $\sim$\arcsa{0}{75}, and the size of the knots is
$\sim$1$^{\prime\prime}$, so the structures of the knots may not be fully
observed.  Therefore, the apparent \textbf{peaks} of the knots may deviate
to \textbf{the north from the model}, but we are not sure how and to what
extent the missing flux impacts our results.  Observations with more
complete UV coverage are needed to confirm this possibility.

It is also possible that the structures of the knots do deviate to
\textbf{the north from the model}.  In the orbiting jet source model
\citep{2010ApJ...713..731L}, the orbit of the jet source was assumed to be
circular.  However, since this reflection-symmetric wiggle is likely due to
an orbital motion of the jet source in a binary system
\citep{2010ApJ...713..731L}, it is possible that the orbit is not perfectly
circle.  It may be elliptical so that the model may under-estimate or
over-estimate the trajectories at different distance
\citep{2004RMxAA..40...61G}.  Therefore, in the high-resolution observation,
the peak positions of the knots can not be fitted exactly by the model. 
Nonetheless, since the binary separation in this model is $\sim$4.6 AU
\citep{2010ApJ...713..731L}, further observations at higher resolution are
also needed to check the possible existence of the binary system.

\subsection{Backward and Forward Shocks and Sideways Ejection.}

As mentioned in Section 3.5, it takes time (or distance as jet propagates)
to fully form internal shocks in knots and sub-knots, and the corresponding
distances in our case are about {\bf\arcsa{2}{6}} and \arcsa{0}{6}, respectively. 
Therefore, we expect to see shock structures, including backward and forward
shocks and sideways ejection in the knots and sub-knots, after these
distances.  Since sub-knots are not spatially resolved, here we focus on the
knots.

\begin{figure}
\epsscale{0.8}
\plotone{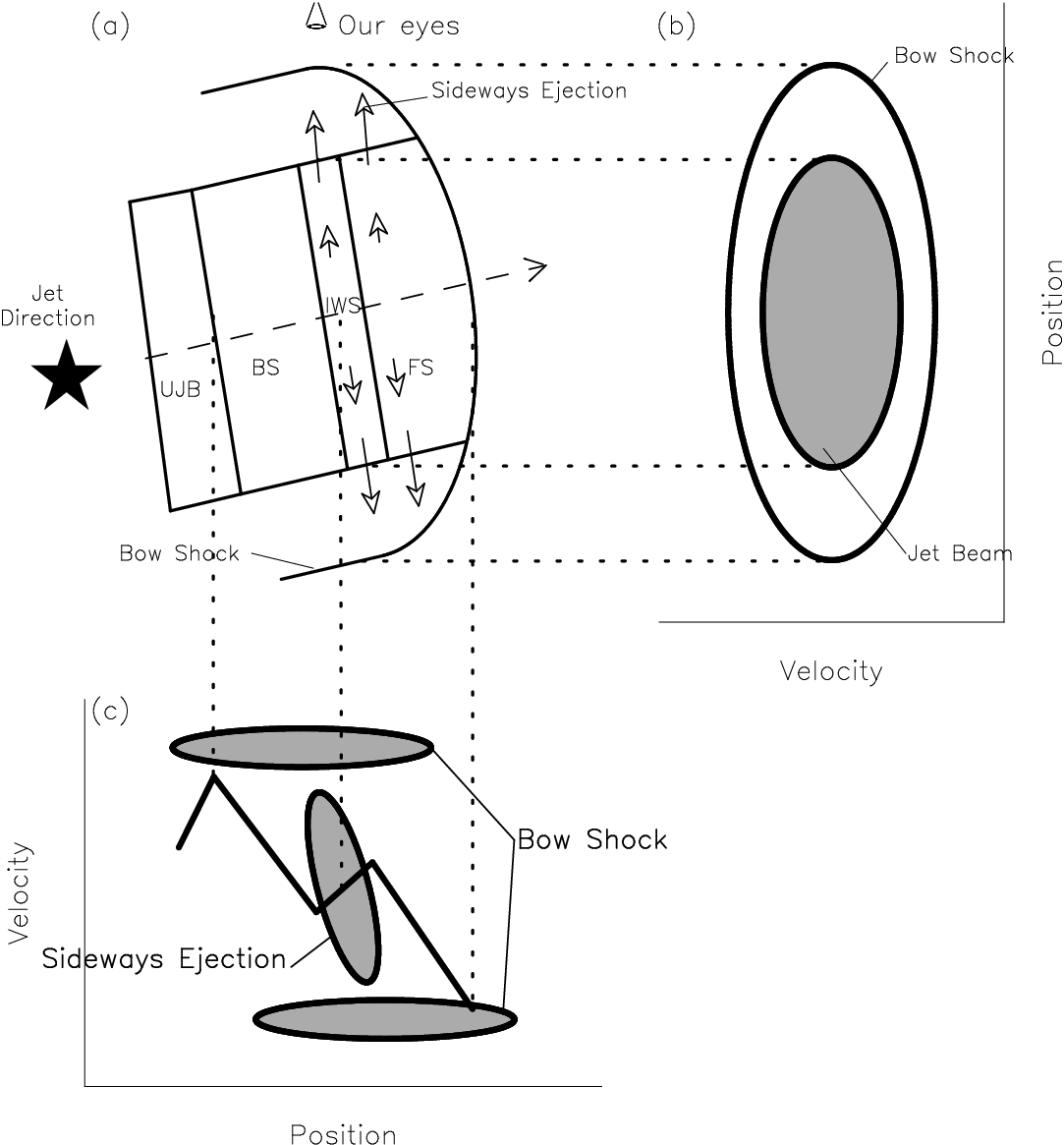}
\centering
\centering
\caption{ (a) Schematic diagram showing the structure of an internal shock, which consists of an unshocked jet beam (UJB), a backward shock (BS), a forward shock (FS), an internal working surface (IWS) and sideways ejection, for a knot. (b) Simplified PV diagram \textbf{along a cut} perpendicular to the jet axis, illustrating how \textbf{the bow shock} would manifest itself in the diagram, see also Figure 26 in \citet{2000ApJ...542..925L}. (c) Simplified PV diagram cut along the jet axis, illustrating how the velocity structure in the knot changes with position, see also Figures 3 and 4 in \citet{1990ApJ...364..601R}, Figure 2 in \citet{1997A&A...318..595S}, and Figure 8 in \citet{2004ApJ...606..483L}.
}
\label{knotstr}
\end{figure}

Figure \ref{knotstr} summarizes the forward and backward shocks, internal
working surface, sideways ejection, and their PV structures in schematic
diagrams
\citep{1990ApJ...364..601R,1993ApJ...413..210S,1997A&A...318..595S,2004ApJ...606..483L,2016ApJ...816...32J}. 
When we use the cut perpendicular to the jet axis (Figure \ref{knotstr}(b)),
\textbf{we can see a hollow elliptical PV structure due to the expanding bow shock
structure.  At the center of this elliptical PV structure, we
can see a filled elliptical PV structure arising from the jet beam with
sideways ejection.} In the cut along the jet axis (Figure \ref{knotstr}(c)),
we can see a zig-zag PV structure containing a forward shock (low velocity),
a backward shock (high velocity), an internal working surface (in the
middle) \citep{2004ApJ...606..483L}, and an elliptical PV structure for the
sideways ejection extents in the middle.

%For sub-knots, in Section 3.4, there are three PV components (see red boxes in Figure \ref{pvbkk} and \ref{pvrkk}). The middle velocity component could trace the expanding shells or sideways ejection of the sub-knots. The triangular shapes are the combinations of circular shapes and a little high-velocity components. However, the resolution is too low to resolve the backward and forward shocks of the sub-knots.

%Then, we will present the study of fully formed knots. As we can see in Figure \ref{siocoso}, knot BK3 are detected in all SiO, CO, and SO. The distance of the peak position of knot BK3 is $\sim$7$^{\prime\prime}$ to the central source, and the size of the knot is $\sim$1$^{\prime\prime}$. As presented in Figure \ref{pvall}(b), the PV structures cut along the jet axis contain a forward shock (low velocity), a backward shock (high velocity), and an internal working surface (in the middle) \citep{2004ApJ...606..483L}. 

\begin{figure}
\epsscale{0.8}
\plotone{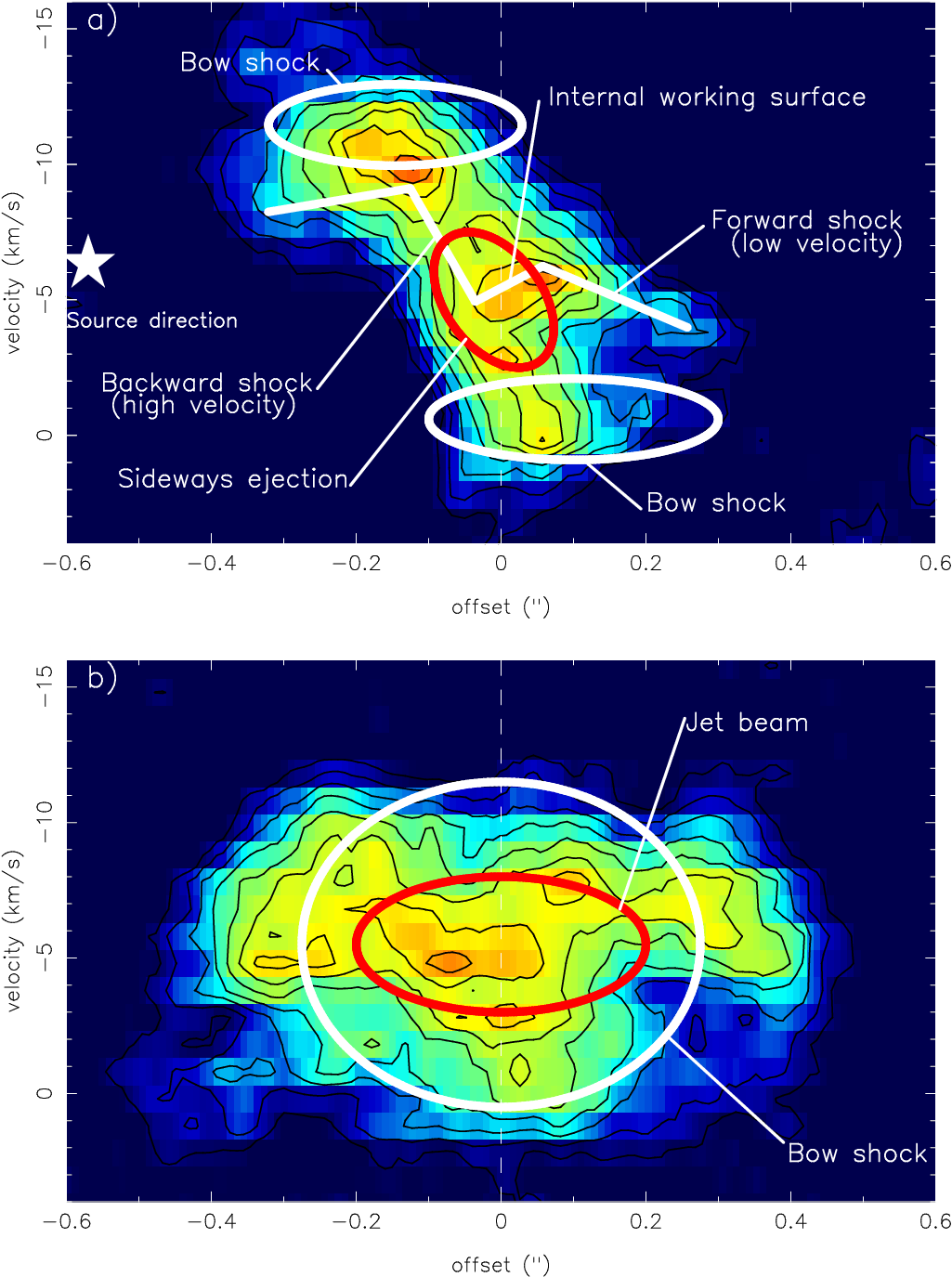}
\centering
\centering
\caption{ (a) SiO PV diagram of knot BK3 cut along the jet axis. The contour levels start at 3 $\sigma$ with a step of 3 $\sigma$, and $\sigma$ is $\sim$0.95 mJy beam$^{-1}$. (b) SiO PV diagram of knot BK3 cut perpendicular to the jet axis. The outer white circle denotes the contribution of sideways ejection, and the inner white circle denotes the contribution of the jet beam.
}
\label{bk32}
\end{figure}

Knot BK3 appears as a bow-like structure and is at a distance much further
away than {\bf\arcsa{2}{6}}, and thus can be used to study the shock structure. 
Figure \ref{bk32}(a) shows a SiO PV diagram of knot BK3 cut along the jet
axis through the peak position.  We can see a zig-zag PV structure
consisting of forward (low velocity) and backward (high velocity) shock and
internal working surface (in the middle), as discussed above.  \textbf{There
are other components in the area of the internal working surface (circled by
a red ellipse and labelled as sideways ejection), and these components may
be due to sideways ejection.} To make it clearer, Figure \ref{bk32}(b) is a
SiO PV diagram of knot BK3 \textbf{along a cut} perpendicular to the jet
axis through the peak position, and it shows an elliptical PV structure
which can be attributed to a {\bf bow shock structure} produced by sideways
ejection.  Comparing Figure \ref{bk32}(a) and (b) (cut through the same
reference central point but different directions), we conclude that the
extra PV component (circled in Figure \ref{bk32}(a)) corresponds to the PV
structure of sideways ejection.  Also, the emission inside the elliptical PV
structure in Figure \ref{bk32}(b) is the PV structure of the internal
working surface (in the jet beam itself), as discussed above.  We also
calculated that the PV structure of sideways ejection along the jet axis
would be $W \cdot sin(i)$ $\sim$\arcsa{0}{2} in extent, where the width of
the knot $W\sim$1$^{\prime\prime}$ and the inclination angle $i\sim$11
degree.  This extent agrees with that seen in Figure \ref{bk32}(a) for the
region labelled as sideways ejection.  Therefore, in our case, the knot can
be resolved to backward and forward shocks plus sideways ejection.  {\bf A
model illustration of these shock structures is given in next section.}

%{\bbf ??? Why discuss Figure 13 before Figure 12??}

Notice that the PV structure of the sideways ejection at different inclination can be different due to the projection effect. For example, the IRAS 04166+2706 jet has a high inclination angle of $\sim$52$^{\circ}$ and the PV structure becomes a sawtooth pattern \citep{2009A&A...495..169S,2017A&A...597A.119T}, whereas HH 212 has a very low inclination angle of $\sim$4$^{\circ}$, and the PV structure becomes arc-like \citep{2015ApJ...805..186L}.

\subsubsection{Illustrative Model For Shock Structures in Knot BK3}

\begin{figure}
\epsscale{0.8}
\plotone{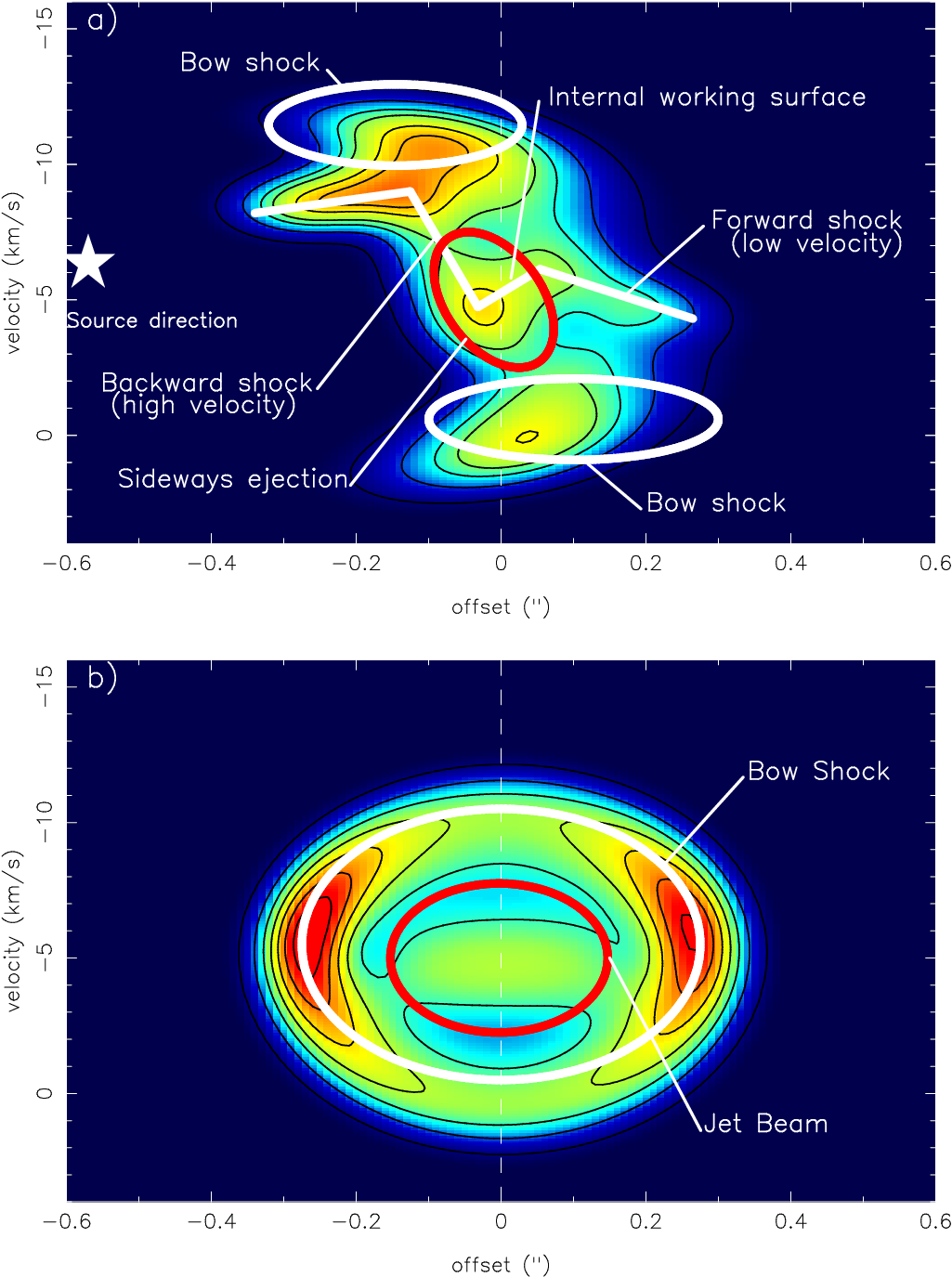}
\centering
\centering
\caption{ \textbf{Same as Figure \ref{bk32} but derived
from a radiative-transfer model discussed in Appendix I.}
}
\label{model}
\end{figure}

\textbf{Here we presented a simple kinematic model to illustrate the shock
structures in knot BK3 in more details.  As shown in Figure \ref{knotstr},
this model has two components, a jet beam and a bow shock structure.  This
jet beam consists of an unshocked jet beam (UJB), a backward shock (BS), an
internal working surface (IWS), and a forward shock (FS), mimicking an
internal shock produced in pulsed jet simulations
\citep{1990ApJ...364..601R,1997A&A...318..595S,2004ApJ...606..483L}.  The
velocity structure in the jet beam has two components; one is along the jet
axis, and the other is perpendicular to the jet axis for the sideways
ejection (see Figure \ref{test}).  The velocity gradients along the jet axis
are so assumed to roughly match the observation.  These velocity gradients
are also expected for an internal shock produced in the pulsed jet
simulations
\citep{1990ApJ...364..601R,1997A&A...318..595S,2004ApJ...606..483L}.  On the
other hand, the velocity structure in the bow shock structure (see Figure
\ref{bow}) is adopted from the jet-driven bow shock model
\citep{2001ApJ...557..443O}.  The resulting PV diagrams are presented in
Figure \ref{model}.  In the PV diagram cut along the jet axis (Figure
\ref{model}a), we see four linear structures as marked by the white lines,
arising from the unshocked jet beam, the backward shock, the internal
working surface, and the forward shock, respectively, roughly consistent
with the observation.  The sideways ejections are mostly from the internal
working surface, causing a spread of velocity there, as marked by the red
ellipse in Figure \ref{model}a.  The PV structures marked by the white
ellipses are from the bow shock structure, with the upper (more blueshifted)
one from the near-side and the lower (less blueshifted) one from the
far-side, also roughly similar to those seen in the observation.  Their
intensity peaks are also at roughly the same locations as those seen in the
observations.  In the PV diagram along a cut perpendicular to the jet axis
(Figure \ref{model}b), there are two components: one is an elliptical PV
structure from the expanding bow shock structure, and the other is a filled
ellipse (marked by the red ellipse) from the jet beam that has a sideways
ejection.  These features are also similar to those seen in the observation.}

\subsection{Relationship between SiO, CO, and SO layers}

%SiO, CO, and SO may be launched from different parts of the protoplanetary disks and form shells or layers of different molecules. Some models, e.g., disk-wind model, have discussed this phenomenon, which tried to answer the question of angular momentum extraction from protoplanetary disks \citep{2017A&A...607L...6T}. However, there is few evidences to show layers of compounds because of low resolution to distinguish such subtle spatial difference. 

%Now, in our observation, as we can see in Figure \ref{siochamap}, \ref{cochamap}, and \ref{sochamap}, different line emissions trace different part of jet and outflow: SiO mainly traces jet, and CO and SO also trace shell-like structure or outflow, which has wider opening angle, in low velocity (near the systemic velocity). Therefore, SiO layer is surrounded by CO and SO layers. However, we still can not separate CO and SO layers.

\begin{figure}
\epsscale{1}
\plotone{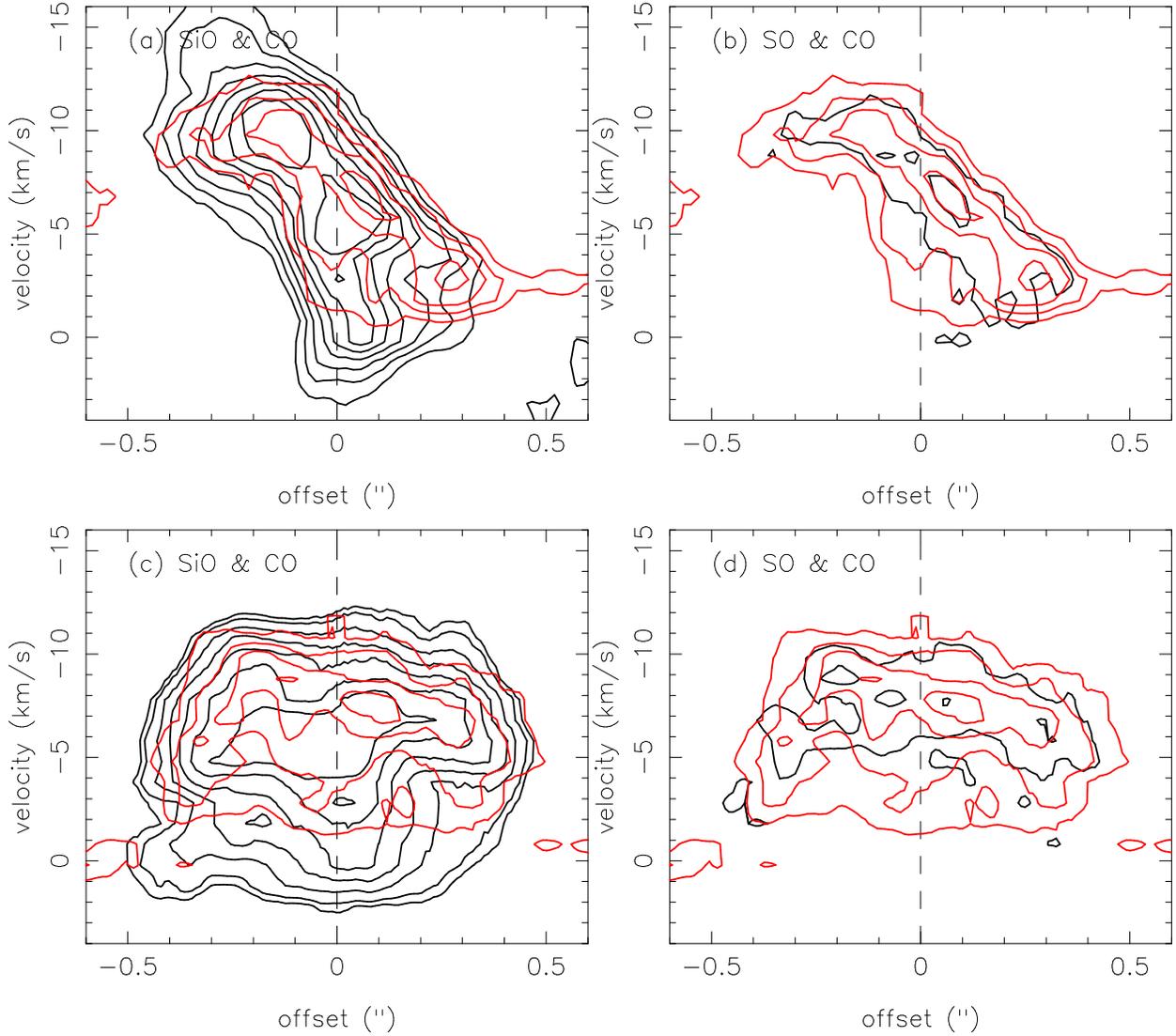}
\centering
\caption{ (a) SiO (black contour) and CO (red contour) PV diagram of knot BK3 cut along the jet axis. For SiO, the contour levels start at 3 $\sigma$ with a step of 3 $\sigma$, and $\sigma$ is $\sim$0.95 mJy beam$^{-1}$. For CO, the contour levels start at 3 $\sigma$ with a step of 3 $\sigma$, and $\sigma$ is $\sim$1 mJy beam$^{-1}$. (b) SO (black contour) and CO (red contour) PV diagram of knot BK3 cut along the jet axis. For SO, the contour levels start at 3 $\sigma$ with a step of 3 $\sigma$, and $\sigma$ is $\sim$1 mJy beam$^{-1}$. (c) SiO (black contour) and CO (red contour) PV diagram of knot BK3 \textbf{along a cut} perpendicular to the jet axis. (d) SO (black contour) and CO (red contour) PV diagram of knot BK3 \textbf{along a cut} perpendicular to the jet axis.
}
\label{bk31}
\end{figure}

\begin{figure}
\epsscale{0.8}
\plotone{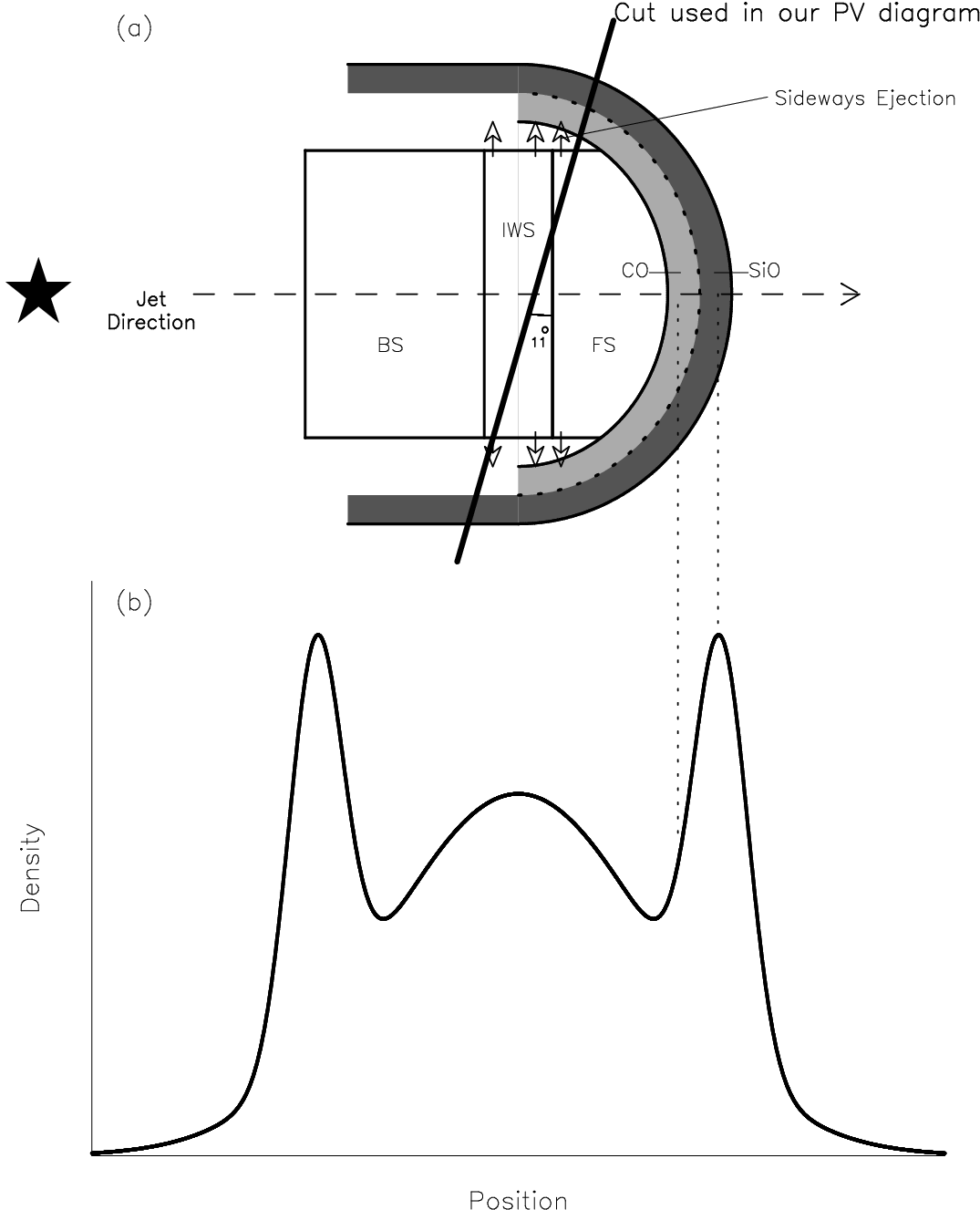}
\centering
\centering
\caption{ (a) Schematic diagram showing the detailed structures of sideways ejection and SiO (dark grey) and CO (light grey) layers. (b) Simplified position to number density diagram cut along the jet axis, illustrating how the density changes with position, see also Figure 2 in \citet{1997A&A...318..595S}, Figure 8 in \citet{2004ApJ...606..483L}, and Figure 7 in \citet{2015ApJ...805..186L}.
}
\label{siocostr}
\end{figure}

Figure \ref{bk31} shows the PV diagrams in SiO, CO, and SO of the fully
formed knot BK3.  CO and SO PV structures are roughly the same, as shown in
Figure \ref{bk31}(b) and (d).  However, the PV structures of CO emission in
Figure \ref{bk31}(a) and (c) only show a part of that of the SiO emission. 
These missing parts of CO PV structures could be due to different emission
lines tracing different layers.  According to Figure \ref{bk31}(c), PV
structures of CO emission do not show the redshifted part of elliptical PV
structures (due to the expanding shell produced by the sideways ejection). 
This could be explained if the SiO layer in the bow shock (as shown in the
dark-grey layer in Figure \ref{siocostr}(a)) extends further away than the
CO layer (as shown in the light-grey layer in Figure \ref{siocostr}(a))
does.  Therefore, the cut inclined to the jet axis by 79 (=90-11) degree
using in our PV diagrams does not go through the expanding shells of CO (see
the dotted curve in Figure \ref{siocostr}(a)) on the far side of the knot
(closer to the central source).  This phenomenon may be due to these two
lines tracing different densities.  The critical density of the SiO(J=8-7)
transition is $\sim$10$^{8}$ cm$^{-3}$ \citep{2009ApJ...699.1584L}, and that
of CO(J=3-2) is $\sim$10$^{5}$ cm$^{-3}$.  So the CO may trace the
low-density part of the bow shock (Figure \ref{siocostr}(b)), and the SiO
layer traces the higher density and outer bow shock (see the dark-grey layer
in Figure \ref{siocostr}(a)).  This phenomenon also appears in HH 212
\citep{2015ApJ...805..186L}, however, we do not see a clear extension of SiO
and CO in our emission maps (higher-resolution observations would be needed
to distinguish it).

%In summary, CO and SO trace the out-most parts of outflow shell, and then all SiO, CO, and SO trace the jet beam. In bow shocks, the SiO traces the outer parts of the bow shock structures, and CO and SO trace the relatively low-density layers covered by the bow shocks. However, at the level of our resolution, we can not distinguish the difference of CO and SO layers.

\section{Conclusions}

We have studied the properties of the HH 211 jet in the CO(J=3-2), SiO(J=8-7), and SO(N$_{J}$=8$_{9}$-7$_{8}$) lines at high resolution with ALMA. Our conclusions are as follows;

1. SiO traces the jet. CO and SO trace the jet at high velocity and outflow shell at low velocity.

2. With our ALMA and 2008 SMA observations, we measured the proper motion of 8 knots, 4 on each side in the inner jet. The proper motion of these knots is roughly the same ($\sim$102 km s$^{-1}$), and consistent with previous measurements in the outer jet.

3. In our high resolution observation, we still see the reflection-symmetric wiggle of the jet, and the orbiting jet source model still fits the trend of the wiggle. However, the peaks of the knots show some deviations from the model.

%4. The PV diagram \textbf{along a cut} perpendicular to the jet axis of the first redshifted knot (knot RK0) shows a {\bf possible} velocity gradient that could be due to rotation in the jet. 

4. The PV diagrams obtained along the jet axis indicates that the jet ejects material continuously with two small periodical variations in velocity. The sub-knot structures are produced by a velocity variation with a shorter period of the two. This also explains why there is no SiO emission between \arcsa{0}{1} to \arcsa{0}{5} and -\arcsa{0}{1} to -\arcsa{0}{5}.

5. The PV diagrams of a fully formed knot BK3 shows a backward and a forward shock and sideways ejection. In knot BK3, the SiO traces the high-density outer layer of the bow shock, while CO and SO trace the low-density inner layers of the bow shock.

\acknowledgments
\textbf{
This paper makes use of the following ALMA data: ADS/JAO.ALMA\#2015.1.00024.S, ADS/JAO.ALMA\#2016.1.00017.S. ALMA is a partnership of ESO (representing its member states), NSF (USA) and NINS (Japan), together with NRC (Canada), MOST and ASIAA (Taiwan), and KASI (Republic of Korea), in cooperation with the Republic of Chile. The Joint ALMA Observatory is operated by ESO, AUI/NRAO and NAOJ.}
We acknowledge grants from the Ministry of Science and Technology of Taiwan (MoST 107-2119-M-001-040-MY3) and the Academia Sinica (Investigator Award AS-IA-108-M01). We also thank Anthony Moraghan for useful discussion. 

\appendix

\section{Appendix I: Shock Model for knot BK3}

\begin{figure}
\epsscale{0.8}
\plotone{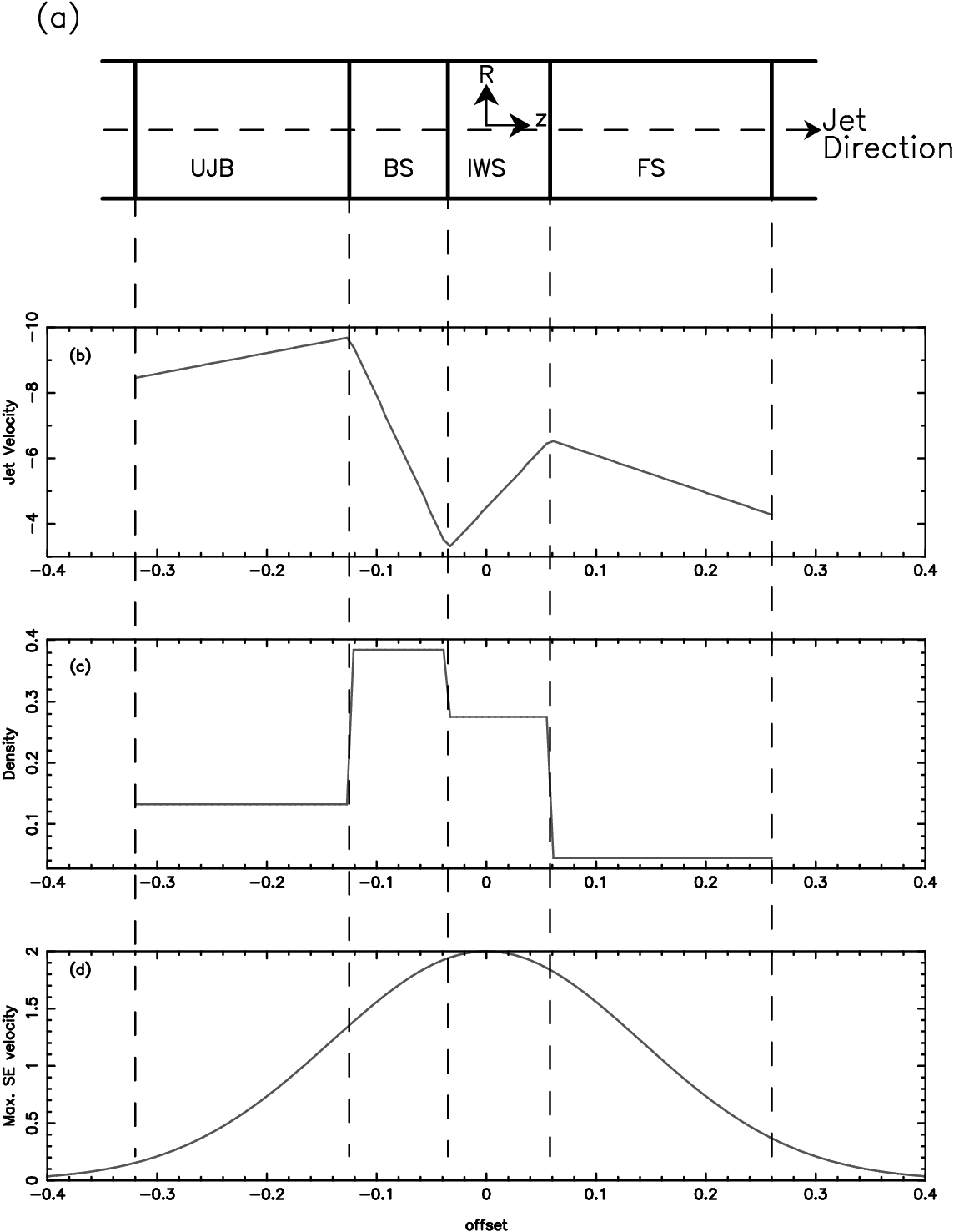}
\centering
\centering
\caption{ (a) Schematic diagram showing the detailed structures of jet beam, 
where UJB, BS, IWS, and FS stand for the unshocked jet beam, 
the backward shock, the internal working surface, and the forward shock, 
respectively. (b) Jet velocity profile along the jet axis. 
(c) Density profile along the jet axis. (d) Maximum 
sideways ejection velocity profile at the jet boundary.
%{\bbf please use larger fonts.}
}
\label{test}
\end{figure}

As mentioned in Section 4.2.1, we constructed a simple kinematic model to
produce the features seen in the observations.  We adopted a cylindrical
coordinate system, and assumed that the jet propagates along the z axis (see
Figure \ref{test}(a)) and has a cylindrical radius R$_{j}$=\arcsa{0}{2}.

The jet beam has four components: UJB (unshocked jet beam), BS (backward
shock), IWS (internal working surface), and FS (forward shock), with their
velocity and density structures in the z axis shown in Figures \ref{test}(b)
and (c), respectively.  The velocity gradients here follow those expected
from the internal shock produced by a periodical variation in the jet
velocity
\citep{1990ApJ...364..601R,1997A&A...318..595S,2004ApJ...606..483L}.

%\begin{center}
%d$_{j}$ = d$_{FS}$ + d$_{IWS}$ + d$_{BS1}$ + d$_{BS2}$

%v$_{j}$ = v$_{FS}$ + v$_{IWS}$ + v$_{BS1}$ + v$_{BS2}$ + v$_{SE}$

%\end{center}

%Note that the velocity are varied linearly and continuously with z axis (along the jet axis) according to the observation (check Figure \ref{bk32}(a) and also Figure \ref{knotstr}(c)).

The jet beam also has a sideways ejection (SE) perpendicular to the z axis.
The sideways ejection is assumed to increase along the R axis from zero to
a maximum value at the jet boundary.  The maximum value at the jet
boundary is assumed to peak at the internal working surface and decreases
toward the backward and the forward shocks (Figure \ref{test}(d)).

%For \arcsa{0}{26} $\ge$ z $>$ \arcsa{0}{058}

%For \arcsa{0}{058} $\ge$ z $>$ -\arcsa{0}{035}

%For -\arcsa{0}{035} $\ge$ z $>$ -\arcsa{0}{125}

%For -\arcsa{0}{125} $\ge$ z $\ge$ -\arcsa{0}{32}

%The velocity structures are present in Figure \ref{knotstr}(c):
%\begin{center}

%V$_{jz}$ = V$_{o}$z + V$_{j}$

%V$_{jr}$ = V$_{ro}$e$^{-z^{2}/z_{4}}$r / r$_{j}$

%\end{center}

\begin{figure}
\epsscale{0.8}
\plotone{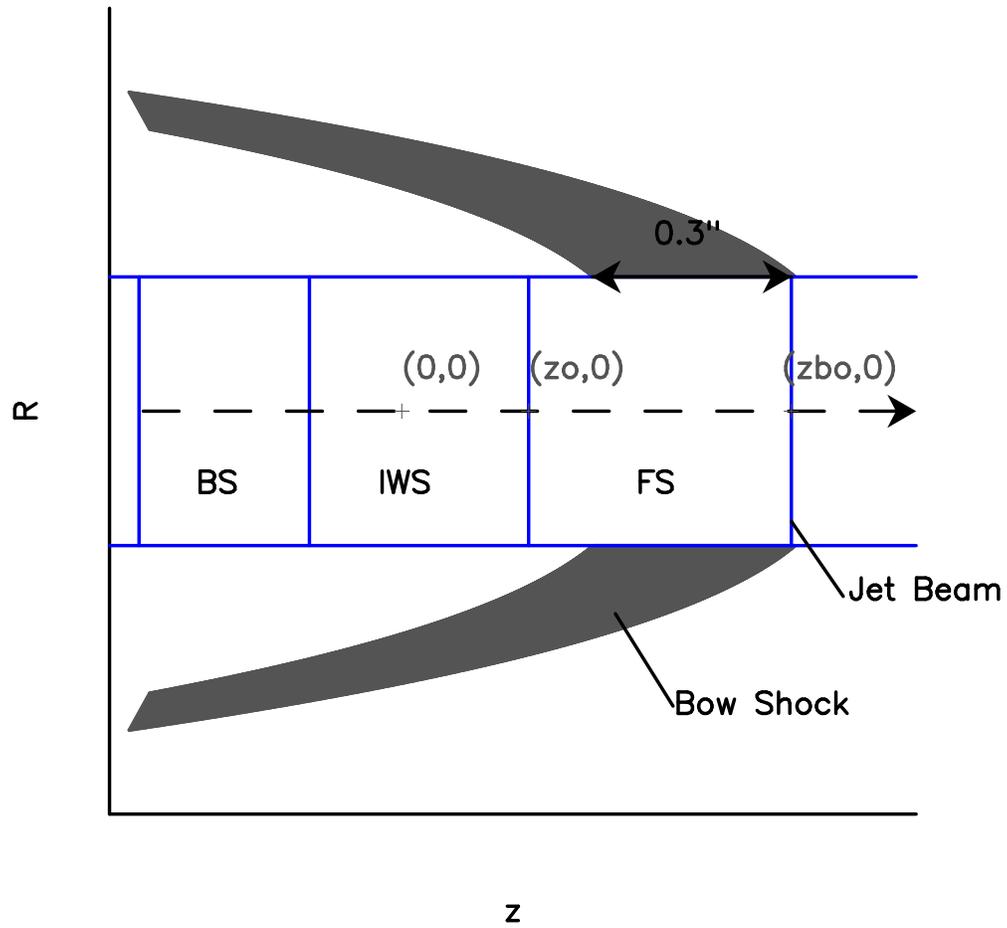}
\centering
\centering
\caption{ Schematic diagram showing the detailed structures of the jet beam and the bow shock, where BS, IWS, and FS stand for the backward shock, the internal working surface, and the forward shock, respectively.
%{\bbf please shade the bow shock structure??}
}
\label{bow}
\end{figure}

For the bow shock around the jet beam (see Figure \ref{bow}), we followed the jet-driven bow shock model reported in \citet{2001ApJ...557..443O}.
The shape of the bow shock is given by:

\begin{center}

z + z$_{bo}$ = -(R/R$_{j}$)$^{3}$(v$_{j}$R$_{j}$) / (3$\beta$c$_{s}$)

\end{center}
, where z$_{bo}=$\arcsa{0}{5} is z position of the bow shock tip (see Figure
\ref{bow}), c$_{s}=8$ km s$^{-1}$ is the isothermal sound speed in the
shock, and $\beta=4$ is to account for the effective impulse from the shock, as
found in \citet{2001ApJ...557..443O}.

The velocity structures of the bow shock structure are:

\begin{center}

V$_{b,z}$ = $\frac{(\beta c_{s} R_{j}^{2})^{2}}{(\beta c_{s} R_{j}^{2})^{2} + v_{s}(R^{2}-R_{j}^{2})^{2}}$ v$_{s}$

V$_{b,R}$ = $\frac{(\beta c_{s} R_{j}^{2})*(R^{2}-R_{j}^{2})}{(\beta c_{s} R_{j}^{2})^{2} + v_{s}(R^{2}-R_{j}^{2})^{2}}$ v$_{s}$

\end{center}
, where v$_{s}=14$ km s$^{-1}$ is the shock velocity to roughly match the observation.

%The values of above constants and functions in the model are taken from the observation and  \citet{2001ApJ...557..443O}.

The bow shock structure has a thickness of \arcsa{0}{3}. Its density
is assumed to be 
\begin{center}

d$_{b}$ =  d$_{o}$e$^{-z^{2}/z_{o}}$

\end{center}
, where z$_{o}$ is a constant. With this assumption, the density is
highest at $z=0$, which is a location aligned with the internal working surface, 
to match the observation.

%Finally, we judged from the observation (Figure \ref{bk32}), then have D$_{o}$ = 1.1x10$^{-1}$ cm$^{-3}$, z$_{1}$ = \arcsa{0}{18},  z$_{2}$ = \arcsa{0}{1}, z$_{3}$ = \arcsa{0}{08}, r$_{1}$ = r$_{j}$ = \arcsa{0}{2}, V$_{o}$ = -33 km s$^{-1}$arcsec$^{-1}$ and V$_{j}$ = 80.675 km s$^{-1}$ at z $\le$ \arcsa{0}{3} and z $\ge$ \arcsa{0}{075}, V$_{o}$ = 16 km s$^{-1}$arcsec$^{-1}$ and V$_{j}$ = 77 km s$^{-1}$ at z $\le$ \arcsa{0}{075} and z $\ge$ -\arcsa{0}{075}, V$_{o}$ = -70 km s$^{-1}$arcsec$^{-1}$ and V$_{j}$ = 70.55 km s$^{-1}$ at z $\le$ -\arcsa{0}{075} and z $\ge$ -\arcsa{0}{2}, V$_{o}$ = 33 km s$^{-1}$arcsec$^{-1}$ and V$_{j}$ = 91.15 km s$^{-1}$ at z $\le$ -\arcsa{0}{2} and z $\ge$ -\arcsa{0}{5}, V$_{ro}$ = 4 km s$^{-1}$, z$_{4}$ = \arcsa{0}{05}, v$_{j}$ = 100 km s$^{-1}$, $\beta$ = 4, c$_{s}$ = 10 km s$^{-1}$, z$_{5}$ = \arcsa{0}{05}, V$_{s}$ = 20 km s$^{-1}$.

\begin{table}
\begin{center}
\caption{Observation Logs}
\begin{tabular}{cccccc}
\hline
Observation & Projected & Angular & Time on & Number of & Sensitivity \\
Time (year) & Baselines (m) & Resolution (SiO) & Target (min) & Antennas & (SiO, mJy beam$^{-1}$) \\\hline
2015.9 & 15-7930 & \arcsa{0}{07} $\times$ \arcsa{0}{05} & 123 & 37-41 & 2.2\\\hline
2016.8 & 13-3145 & \arcsa{0}{13} $\times$ \arcsa{0}{08} & 132 & 43 & 1.0\\\hline
\end{tabular}
\end{center}
\end{table}

%\begin{figure}
%\epsscale{0.8}
%\plotone{SiO_RK1_h.eps}
%%\includegraphics[scale=1]{SiO_RK1_h.png}
%\centering
%\caption{ Blueshifted and redshifted SiO emission of the jet. For knot BK0, the velocity ranges of the blue- and red-shifted emission are $\sim$ -14.8 to -8.8 km s$^{-1}$ LSR and -7.8 to 1.2 km s$^{-1}$ LSR, respectively. For knot RK0, the velocity ranges of the blue- and red-shifted emission are $\sim$ 24.2 to 31.2 km s$^{-1}$ LSR and 32.2 to 41.2 km s$^{-1}$ LSR, respectively. The contour levels start at 3 $\sigma$ with a step of 3 $\sigma$, and $\sigma$ is $\sim$0.0009 Jy beam$^{-1}$.
%{\bbf Please replace this Figure with new one...}
%}

%\label{rk1h}
%\end{figure}

%\begin{figure}[tbh]
%\includegraphics[scale=0.8]{pgplot.png}
%\caption{\em{The CO(1-0) velocity field of NGC\,3256, with contours 
%of the total line emission map overlaid (ALMA Science Verification Data).
%}}
%\end{figure}

\clearpage

\end{document}